\documentclass[10pt,journal,compsoc]{IEEEtran}
\usepackage{amssymb}
\usepackage{amsmath}
\usepackage{multicol}
\usepackage{subcaption}
\usepackage{multirow}
\usepackage{algorithm}
\usepackage{algorithmic}
\usepackage{url}
\usepackage{pifont}
\usepackage{float}
\usepackage{graphicx}
\usepackage[table,xcdraw]{xcolor}
\newtheorem{theorem}{Lemma}

\ifCLASSOPTIONcompsoc
  \usepackage[nocompress]{cite}
\else
  \usepackage{cite}
\fi
\ifCLASSINFOpdf
\else
\fi
\begin{document}
%
\title{Federated Route Leak Detection in Inter-domain Routing with Privacy Guarantee}

%
%
%
%

\author{Man~Zeng,~Dandan~Li,
        ~Pei~Zhang,
        ~Kun~Xie,~Xiaohong~Huang 
        \thanks{Corresponding author: Xiaohong Huang, \emph{huangxh@bupt.edu.cn} }
        
  }
\ifCLASSOPTIONpeerreview
  \markboth{Journal of \LaTeX\ Class Files,~Vol.~14, No.~8, August~2015}%
  {Shell \MakeLowercase{\textit{et al.}}: Bare Demo of IEEEtran.cls for Computer Society Journals}
\fi

%



\IEEEtitleabstractindextext{%
\begin{abstract}
  In the inter-domain network, a route leak occurs when a routing announcement is propagated outside of its intended scope, which is a violation of the agreed routing policy. The route leaks can disrupt the internet traffic and cause large outages. The accurately detection of route leaks requires the share of AS business relationship information of ASes. However, the business relationship information between ASes is confidential due to economic issues. Thus, ASes are usually unwilling to revealing this information to the other ASes, especially their competitors. Recent advancements in federated learning make it possible to share data while maintaining privacy. Motivated by this, in this paper we study the route leak problem by considering the privacy of business relationships between ASes, and propose a method for route leak detection with privacy guarantee by using blockchain-based federated learning framework, in which ASes can train a global detection model without revealing their business relationships directly. Moreover, the proposed method provides a self-validation scheme by labeling AS triples with local routing policies, which mitigates route leaks' lack of ground truth. We evaluate the proposed method under a variety of datasets including unbalanced and balanced datasets. The different deployment strategies of the proposed method under different topologies are also examined. The results show that the proposed method has a better performance in detecting route leaks than a single AS detection regardless of whether using balanced or unbalanced datasets. In the analysis of the deployment, the results show that ASes with more peers have more possible route leaks and can contribute more on the detection of route leaks with the proposed method.
\end{abstract}

}

\maketitle

\IEEEdisplaynontitleabstractindextext

%
\IEEEpeerreviewmaketitle


\section{Introduction}\label{sec:introduction}
Border Gateway Protocol (BGP) is used in the inter-domain network for exchange of routing information between autonomous systems (ASes). In BGP, each AS selects the best route according to its routing policies and announces the selected route to neighbors.  Different from the shortest path routing policy in the intra-domain, the routing policy in the interdomain network is more complicated since it considers business relationships between ASes, where the business relationship can be categorized into two types according to AS economic factors \cite{luckie2013relationships}: \textit{customer-to-provider (c2p)} and \textit{peer-to-peer (p2p)}. 

 The inter-domain routing policies have been extensively investigated in a number of studies, such as \cite{gao2001inferring,gill2013survey,anwar2015investigating}. Their analysis suggests that a policy most commonly adopted is \textit{valley-free rule} \cite{li2015route}. In the valley-free rule, routes learned from providers or peers should not be exported to other providers or peers.
However, due to misconfiguration and malicious attacks, the routing announcement may be propagated in violation of their agreed routing policy, which is defined as \textit{route leak} \cite{sriram2016problem,galmes2020preventing}.

Route leaks can cause major outages by redirecting traffic and bring a risk of encountering Man-in-the Middle attacks \cite{hepner2009defending}. For instance, during March 12, 2015, a broadband provider of India (AS17488) wrongly announced over three hundred Google's prefixes to its provider AS9498, making many of Google's services inaccessible to their users \cite{routleak2015google}. Another incident occurred on February 11, 2021, AS28548 in Mexico leaked more than two thousand prefixes to its neighbors and affected about 80 countries in the world \cite{exampleRouteLeak}. The detection of route leak is becoming increasingly important, given the growing number of serious route leak reports.

The major challenge of detecting route leaks is that the business relationships of ASes are confidential.
Each AS only knows the relationships between itself and its neighbors but does not know the exact relationship of others due to privacy issues. In order to detect route leaks, some studies \cite{luckie2013relationships,jin2019stable, jin2020toposcope,shapira2020unveiling} focus on inferring AS business relationships.
However, their inference techniques still suffer errors on partial critical links \cite{jin2020toposcope}.  Studies like \cite{galmes2020preventing,he2020roachain,chen2020isrchain,sriram2017methods} do not consider the privacy of AS relationship well and require deployers directly revealing AS relationship information, which makes the deployment hard to proceed.

Another challenge for route leak detection is the lack of ground truth. Only a few destructive routing leak events reported have been validated. Route leaks that are not related to the customer-aware services, e.g. multi-media services, are hard to be validated. For example, S. Abd El Monem et al. \cite{abd2020bgp} show there were only 13 validated route leak incidents between 2006 and 2018. The lack of a ground truth makes popular techniques such as traditional machine learning techniques difficult to be utilized in route leak detection.

Federated learning is a distributed machine-learning method that allows participants globally to train a model without needing to transport their local training data to a central server.
Moreover, instead of aggregating local training data, federated learning aggregates local model updates of participants, which can protect data privacy of participants. However, traditional federated learning methods require a third-party server for aggregating model updates, which is vulnerable to single point of failure. To avoid this problem, blockchain-based federated learning framework \cite{hou2021systematic,shayan2020biscotti,majeed2019flchain,lu2019blockchain} is proposed, 
where blockchain can provide security management of participants, auditability, and so on.

In this paper, we propose a method to route leak detection using a blockchain-based federated learning framework. As outlined previously, one AS has limited AS business relationship information, while the detection of route leaks needs as much relationship information as possible. 
By using federated learning, ASes can globally train a model to identify route leaks with sharing model updates instead of directly sharing their business relationships with others. Furthermore, in order to further strengthen the privacy protection of relationship information, we propose to replace business relationships with AS triples to train the models. By  labeling AS triples as malicious or regular using local routing policies, our method provides a solution to solve the lack of ground truth in route leaks. 

In the proposed method, AS participants firstly use their known routing policies to generate local training data for training. The local training data are composed of AS triples and corresponding labels that show the triples are malicious or regular. Then, each AS participant uses its local training data to train a local model and exchange updates of the local model with each other through the blockchain network. In each global communication round, one AS can gather all local model updates of participants and aggregate these updates. After the training is finished, the final global model updates are stored in the blockchain. AS participants can retrieve the global model from blockchain and utilize it for further route leak detection.

The contributions of this paper are summarized as follows:
\begin{itemize}
    \item We consider the problem of route leak detection and propose a privacy-preserving method for sharing routing policy information in AS triples compared to methods that directly share business relationships among ASes. 
    \item We customize a blockchain-based federated learning framework to learn routing policies in inter-domain networks and globally train a model to detect route leaks accurately. 
    \item We evaluate the proposed method on different datasets and analyze different deployment strategies of the method under different topologies. Results show that the proposed method can improve the performance of a single AS in detecting route leaks.
    The results also indicate that AS with more peers appears to have more possible route leaks and can contribute more to the detection of route leaks than others if they deploy the method.
\end{itemize}

The remainder of the paper is organized as follows. Section \ref{sec:backgroud_related} gives an overview of route leak definitions in inter-domain networks and related works in detecting route leaks. In Section \ref{sec:method}, we introduce the route leak detection problem and the proposed solution method. Section \ref{sec:experiments} introduces the conducted experiments and the analysis of results. Section \ref{sec:conclusion} gives the conclusion of this paper.

\begin{figure*}[htbp]
    \centerline{\includegraphics[scale=0.32]{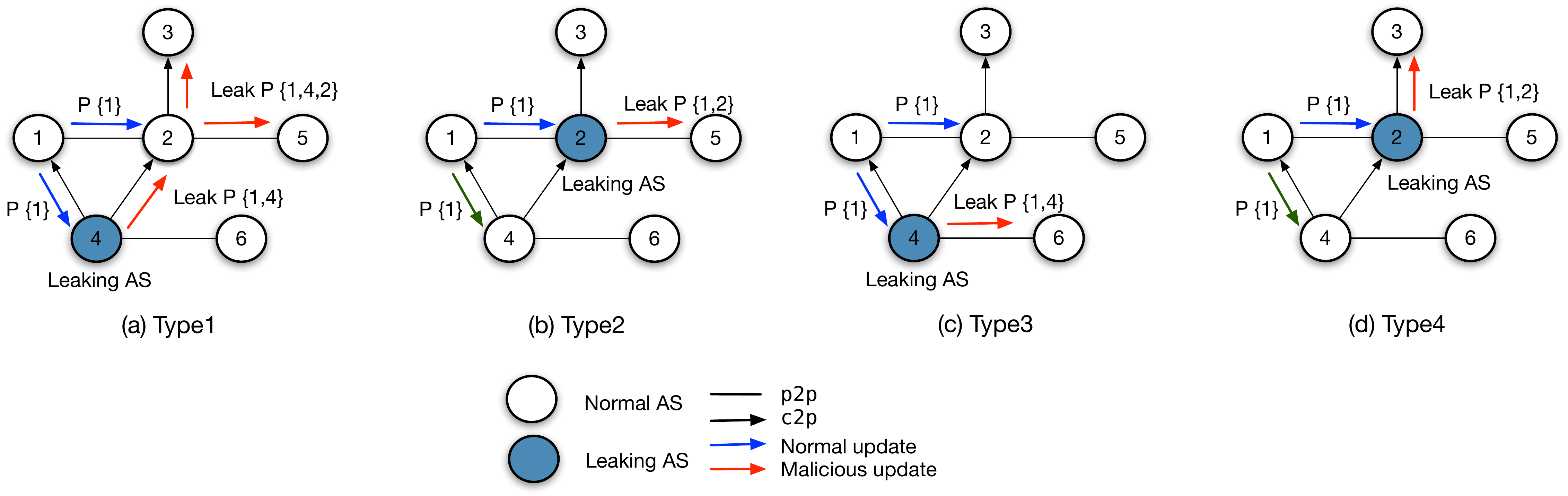}}
    \caption{Route leak examples of Type1 to Type4 route leaks: a) Type1: AS4 leaks route $P$ learned from its provider AS1 to its another provider AS2; b) Type2: AS2 leaks route $P$ learned from its peer AS1 to its another peer AS5; c) Type3: AS4 leaks route $P$ learned from its provider AS1 to its peer AS6; d) Type4: AS2 leaks route $P$ learned from its peer AS1 to its provider AS3. The p2p represents peer-to-peer relationship and c2p represents customer-to-provider relationship}
    \label{fig:routeleaktypes}
\end{figure*}

\section{Background and related works}\label{sec:backgroud_related}
\subsection{Route Leaks}\label{sec:routeleaks}
Route leaks defined in RFC 7908 \cite{sriram2016problem} are categorized into six types: Type1: a multihomed AS (has more than one provider) leaks the routes learned from its one provider to its another provider. Type2: an AS leaks the routes learned from its one peer to its another peer. Type3: an AS leaks learned routes from its provider to its peer. Type4: an AS leaks of the routes learned from its peer to its provider. Type5: a multihomed AS announces the routes learned from its one upstream ISP as its origin routes to its another upstream ISP. Type6: an AS leaks its internal routes to its provider or peer.

The Type5 and Type6 leaks are more related to the origin prefix hijacking and can be detected by Route Origin Validation (ROV) techniques, such as Resource Public Key Infrastructure (RPKI) \cite{lepinski2012rfc}. RPKI can build a trusted repository for validating ownership certificates of prefixes and origin ASN. Type1, Type2, Type3 and Type4 leaks are primarily grouped by ASes business relationships.  Fig.\ref{fig:routeleaktypes} illustrates simple examples of these 4 types of route leaks, where $P$ is the announcing route. In this paper, we mainly focus on the detection of Type1 to Type4 leaks.

\subsection{Route leaks detection}
The Internet Routing Registries (IRRs) are public distributed databases that allow ASes to register and share their routing policies. Using information in IRR, ASes can detect route leaks that violate routing policies. However, IRR databases are managed by different Regional Internet Registries (RIRs) or private organizations \cite{mcdaniel2020flexsealing}, which pose challenges for information synchronization and consistency. 

Route-leak Protection (RLP) \cite{sriram2017methods} adds a new BGP community attribute named Down-only to the BGP update message for detecting and mitigating route leaks. When receiving a new route, ASes will check the value in the Down-only community of the route to decide whether to forward it to upstream providers/peers or not. If the route comes from the customer or peer ASes and the community shows no forwarding to upstream ASes, then the route is marked as route leak. However, the performance of RLP can be easily affected by misconfiguration of BGP communities and maliciously modifications or discarding of Down-only communities during route propagation \cite{jia2study}.

S. Li et al. \cite{li2015route} conclude a relationship between route loops and route leaks from analyzing local routing information and develop an algorithm for detecting route leaks. However, only some route leaks may cause route loops. Therefore, it can only detect part of route leaks. Similar to \cite{li2015route}, M. Siddiqui et al. \cite{siddiqui2015self} also develop a theoretical framework for detecting route leaks. However, the detection method is valid in route leak initiations and is not suitable for detecting route leak propagation \cite{siddiqui2015self}.

Autonomous System Provider Authorization (ASPA) \cite{azimovverification} is based on the RPKI and adds a new type of object to the RPKI repository. The new object includes the pair of downstream AS and authorized upstream AS. The authorized upstream AS is allowed to propagate the downstream AS's announcements.  By validating the certificates of pairs, ASes can detect route leaks. However, it cannot validate complex relationships like mutual transit, where ASes provide transit service to each other \cite{li2015route}.

There are studies \cite{galmes2020preventing,he2020roachain,chen2020isrchain} based on blockchain to prevent route leaks. In the blockchain, ASes share their relationships and store them in blocks. However, it also direct expose confidential AS business relationship, which influences incentive of deployment.
J. Yue et al. \cite{yue2021privacy} consider the privacy of AS policy when using blockchain for route leak detection. They use Trusted Execution Environment (TEE) \cite{sabt2015trusted} to implement the privacy protection. However, it requires each chain node to maintain a global confidential and tamper-proof routing policy repository, which adds the risk of routing policy leakage if there is a chain node that acts maliciously. Moreover, since TEE uses a combination of hardware and software for protecting the data privacy, it is affected by hardware vulnerabilities and its updates require hardware update.

In this paper, we propose a method for route leak detection using blockchain-based federated learning, which considers the privacy protection of AS business relationship information and uses AS triples instead of AS business relationships to train models. The proposed method keeps the data local and shares only model updates, which lowers the risk of data leakage. With the help of blockchain technique, the proposed method can audit and track data. Even if one AS acts maliciously and leaks the updates, it does not leak the direct business relationship information. Besides, since the method uses AS triples, the complex relationships such as mutual transit can also be handled.

\section{Methodology}\label{sec:method}
\begin{figure*}[htbp]
    \begin{center}
        \centerline{\includegraphics[scale=0.25]{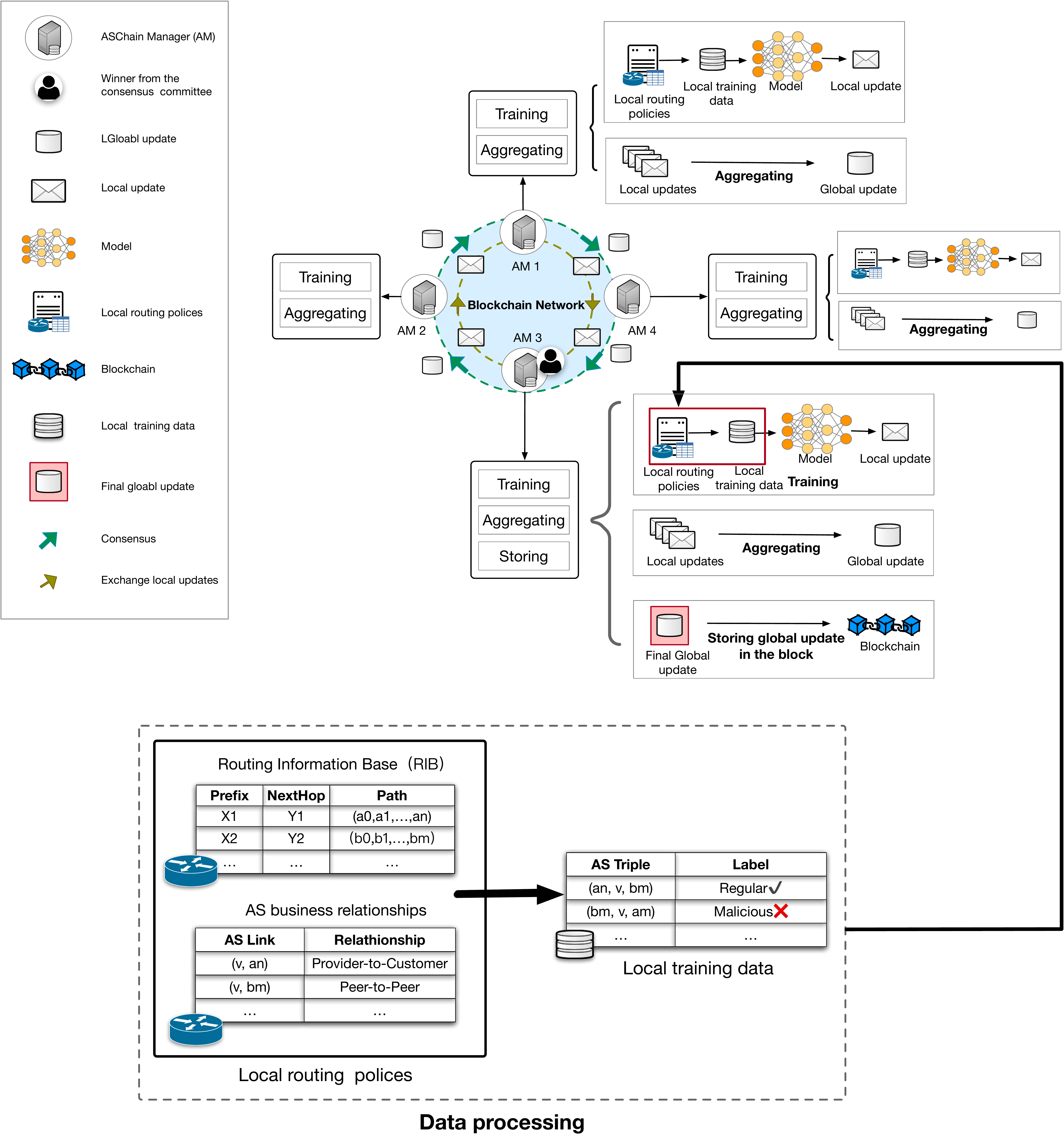}}
       
        \caption{Overview of the solution framework.}
        \label{fig:fl-framework}
        \end{center}
\end{figure*}
In this section, we first give a description about the route leak problem in inter-domain networks (Section \ref{subsec:problem}) and an overview of the proposed framework to solve the problem (Section \ref{subsec:overview}). Following that, we describe the data processing details of how to generate local training data using known local routing policies (Section \ref{subsec:datageneration}). Then, we introduce the model sharing and storing during the blockchain (Section \ref{subsec:sharing}), and discuss the total theoretical cost of a training task (Section \ref{subsec:cost}). Last, an analysis of factors affecting the deployment of the proposed route leak detection method is presented (Section \ref{subsec:factors}).


\subsection{Problem description}\label{subsec:problem}
As introduced in Section \ref{sec:routeleaks}, we can conclude that the route learned from one provider or peer cannot be exported to another provider or peer. For example, as illustrated in Fig.\ref{fig:routeleaktypes}, AS4 leaks the route learned from its provider AS1 to its another provider AS2 due to misconfiguration. The AS2 then forwards the route to its other neighbors. For AS2 and its peer AS5, they only know the business relationship between their direct neighbors and themselves, but they do not know the accurate business relationship between AS1 and AS4, which makes it difficult for both AS2 and AS5 to identify the leak route.
If AS4 shares its relationship with AS1 to AS2 or AS5, at least one of them can detect the route leak and stop propagating the malicious route. In other words, the challenge of route leak detection is to get as accurate AS business relationships of other ASes as possible.

The inter-domain network is a distributed but cooperative system where each ASes exchange routes, which indicates that AS business relationships of an individual AS cannot be hidden completely in the real network \cite{xiang2013sign}. 
Even though some business relationships have already been exposed due to public BGP updates, there is still a hesitation from ASes to announce business relationship information proactively because of business competition. Therefore, in order to facilitate the deployment of detection systems, we consider increasing the difficulty of obtaining information about relationships from the outside to protect privacy.
We do not require ASes to directly expose their business relationships. Instead, we use \textit{AS triple} \cite{li2015route} $(v_{i-1},v_i,v_{i+1})$ with a malicious or regular label to hidden AS business relationships, 
where $v_{i-1}$ and $v_{i+1}$ are both direct neighbors of AS $v_i$. That is, if the triple $(v_{i-1}, v_{i}, v_{i+1})$ violates the routing policy, then it is labeled as malicious. 
Thus, the route leak detection problem is transformed into sharing AS triples with labels as much as possible. 

To solve the problem defined above, we propose a method called \textit{FL-RLD} to share AS triples using blockchain-based federated learning framework, where blockchain is used to provide a secure system for federated learning \cite{zheng2018blockchain}.
In comparison with using a distributed repository to store these AS triples, federated learning provides lower risk of privacy leakage because it allows keeping private data locally and uploading the updates of trained local models instead of AS triples. As a result, we further transform the problem from sharing AS triples into sharing local model updates. 
\begin{figure}[!h]
    \begin{center}
        \centerline{\includegraphics[scale=0.35]{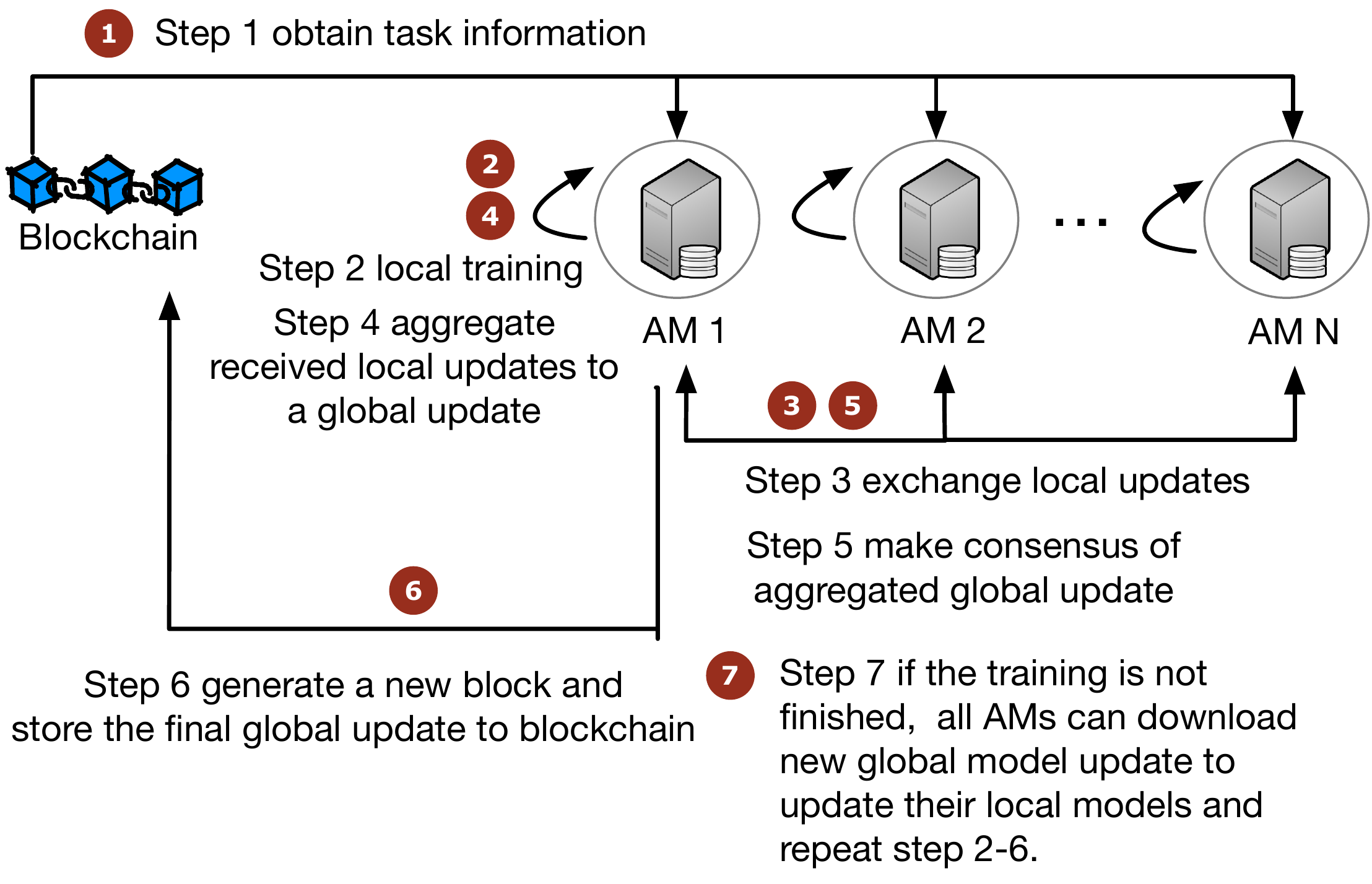}}
       
        \caption{Overview of training process where AM represents ASchain Manager.}
        \label{fig:fl-flow}
        \end{center}
\end{figure}

\subsection{Overview of solution framework }\label{subsec:overview}
The solution framework is shown in Fig.\ref{fig:fl-framework}. 
In the framework, each AS deploys a blockchain node called ASChain Manager. Assuming most ASes are honest but curious. 
ASchain Manager is responsible for managing the blockchain network and federated learning tasks. 
As shown in Fig.\ref{fig:fl-framework}, each ASchain Manager has at least two stages in one task. The first phase is to train its local model with local training data and upload updates of the local model to the blockchain network. The local training data is generated by local routing policies and the details about the generation will be introduced in Section \ref{subsec:datageneration}.
The second phase is to aggregate the received updates to a global model update, and upload the aggregated updates to make consensus. After the consensus, the winner of the consensus committee has the third phase, which is to store the global model update in the blockchain. In light of numerous studies \cite{shayan2020biscotti,li2020blockchain, korkmaz2020chain} have made achievements on the consensus mechanism and other security issues of blockchain-based federated frameworks, we move our focus on customizing the framework for learning inter-domain routing policies. 

The specific training process is illustrated in Fig.\ref{fig:fl-flow}. In the initial state, assuming that an authorized organization publishes a learning task in the blockchain network and each ASchain Manager in the network can obtain all requirements of the learning task, such as initial training model and training epoch. 
(step \ding{172}) When the task starts, ASchain Managers retrieve the task information from the blockchain. (step \ding{173}) ASchain Managers use local training data to train their local models. The local training data is composed of AS triples with labels generated using local routing policies.
(step \ding{174}) In each global training epoch, the ASchain Manager encrypts the update of a local model and propagates it to the network. Other ASchain Manager decrypt and verify the received local update. 
(step \ding{175}) After all local updates are verified, the ASchain Manager begins to aggregate these local updates to a global model update and (step \ding{176}) propagates it to the network to make consensus with others. 
(step \ding{177}) The winner of the consensus generate a new block and store the final global update to blockchain.
(step \ding{178}) If the training is not finished, the ASchain Manager updates the local model using the agreed global model update and repeats the step \ding{173} to \ding{177}.

\begin{algorithm}[htbp]
  \caption{Data Processing}
  \label{alg:dataprocess}
  \begin{algorithmic}[1]
      \STATE \textbf{Input}: deployed AS $m$, RIB $S_m$, neighbors $N_{m}$, business relationship $R_m$.
      \STATE \textbf{Output}: Training data $D_m$ 
      \STATE $D_m \gets \varnothing$
      
      \FOR{$\forall n_i \in N_{m}$}
      \FOR{$\forall n_j \in N_{m}$}
      \IF{$n_i = n_j$}
      \STATE \textbf{Continue}
      \ENDIF
     
      \STATE // \textit{Direct triples} 
      \IF{$R_m(m,n_i)$ = c2p or $R_m(m,n_i)$ = p2p }
      \IF{$R_m(m, n_j)$ = c2p or $R_m(m,n_j)$ = p2p}
      \STATE /* AS triple $(n_i, m, n_j)$ is labeled as malicious */
      \STATE $D_m \gets D_m \cup \{(n_i, m, n_j, 0)\}$
      \ENDIF
      \ENDIF
      \STATE /* AS triple $(n_i, m, n_j)$ is labeled as regular */
      \STATE $D_m \gets D_m \cup \{(n_i, m, n_j, 1)\}$
     
      \STATE // \textit{Inference triples}
      \IF{link $(n_i, n_j)$ in $S_m$ }
      \STATE $D_m \gets D_m \cup \{(n_i, n_j, m, 1)\}$
      \STATE /* Reverse triple pattern */
      \STATE $D_m \gets D_m \cup \{(m, n_j, n_i, 1)\}$
      \ELSE 
      \STATE $D_m \gets D_m \cup \{(n_i, n_j, m, 0)\}$
      \ENDIF
      \ENDFOR
      \ENDFOR
  \end{algorithmic}
\end{algorithm}

\subsection{Data processing}\label{subsec:datageneration}
In the next step, we process the original local data of ASes to training data that consists of AS triples and labels indicating whether they are malicious or not. 
The labels are generated according to local routing policies, which include AS business relationships and stable Routing Information Base (RIB) of AS. The RIB contains the information about the selected routes. The combination of an AS triple and its label is defined as a sample. Therefore, for each AS, the training data can be divided into two components, direct samples based on relationship information and inferred samples based on stable RIB. 

Please note that in practice, in addition to the valley-free rule, other routing strategies can also be used to label AS triples. For instance, for an AS triple $(a,b,c)$ where $a$ is a customer of $b$ and $b$ is a customer of $c$, AS $b$ may not allow the route learned from $a$ to export to AS $c$ even it does not break valley-free rule. However, as the confidentiality of other individual routing policies and common use of the valley-free rule, we mainly consider labeling AS triples based on valley-free rule in the experiment. 

In a direct sample $(v_{i-1}, v_{i}, v_{i+1}, \omega)$ of AS $v_{i}$, the possible leaker is $v_{i}$, and the $v_{i-1}, v_{i+1}$ are both direct neighbors of $v_{i}$. The $\omega$ is the label of direct AS triple $(v_{i-1}, v_{i}, v_{i+1})$.
In a inferred sample $(v_{i+1}, v_{i-1}, v_{i}, \omega)$ of AS $v_{i}$, the possible leaker is $v_{i-1}$, and the $v_{i+1}, v_{i-1}$ are both direct neighbors of $v_{i}$. The $\omega$ is the label of inferred AS triple $(v_{i+1}, v_{i-1}, v_{i})$. 
The direct samples are based on the known business relationship with neighbors. The inferred samples are based on the permutations of known neighbors and are used to broaden the vision of AS $v_i$, because direct samples can only include the scenarios about the one-hop vision of $v_i$ and the inferred samples contain part of two-hop vision. The main difference of the two types of triples is that in the direct triples the possible leaker is AS itself $v_i$ and in the inference triples the possible leaker is the neighbor of $v_i$. Details about the process for labeling AS triples are described below. 
\\~\\
\textbf{Triple Labeling}

\begin{enumerate}[]
    \item For each triple $(v_{i-1}, v_{i}, v_{i+1})$ in the direct triples $DiT$ of $v_{i}$: 
    \begin{enumerate}
        \item If $v_{i}$ is a customer or peer of $v_{i-1}$, and $v_{i+1}$ is the provider or peer of $v_{i}$, then the triple breaks the valley-free rule and the label $\omega$ for $(v_{i-1}, v_{i}, v_{i+1})$ is set as malicious.
        \item Otherwise, the label $\omega$ for $(v_{i-1}, v_{i}, v_{i+1})$ is set as regular.
    \end{enumerate}
    \item For each triple $(v_{i+1}, v_{i-1}, v_{i})$ in the inference triples $InT$ of $v_i$: 
    \begin{enumerate}
        \item If the link $(v_{i+1}, v_{i-1})$ never appears in the local stable RIB of $v_i$, then the label $\omega$ for $(v_{i+1}, v_{i-1}, v_{i})$ is set as malicious.
        \item Otherwise, the label $\omega$ for $(v_{i+1}, v_{i-1}, v_{i})$ is set as regular

    \end{enumerate}
    
\end{enumerate}

After the above generation, each AS can obtain a variety of triples including malicious and regular ones. In order to further enrich the triples for training, we conclude a relation pattern about the labels of triples and labels of their reverse triples, which can be summarized as follows. 
\\~\\
\textbf{Reverse Triple Pattern.} For each triple $(v_{i+1}, v_{i-1}, v_{i})$  in inference triples $InT$ of $v_i$:
\begin{enumerate}[]
    \item If the link $(v_{i+1}, v_{i-1})$ appears in the stable RIB:
    \begin{enumerate}
        \item if the triple's label is regular, then the reverse triple $(v_{i}, v_{i-1}, v_{i+1})$ of $(v_{i+1}, v_{i-1}, v_{i})$ is labeled as regular.
        \item if the triple's label is malicious and $v_{i-1}$ is a peer or customer of $v_{i}$, then the reverse triple $(v_{i}, v_{i-1}, v_{i+1})$ of $(v_{i+1}, v_{i-1}, v_{i})$ is labeled as malicious.
    \end{enumerate}
    \item Otherwise, break.
\end{enumerate}

Using the reverse triple pattern, we extend more triples from the initial generated triples. These triples and their labels are combined as local training data for federated learning. For a clear expression, the use of the \textbf{Triple Labeling} and \textbf{Triple reverse pattern} are summarized in Algorithm \ref{alg:dataprocess}

\subsection{Model sharing and storing during the blockchain}\label{subsec:sharing}
In the solution framework, excepet for avoiding single-point failure in traditional federated learning, the blockchain is used to share and store model updates, such as securely exchanging information between participants and auditing uploaded data.  

\textbf{Model sharing:} For example, AS participants in the blockchain can be authorized a public/private key pair (or a set of key paris) with their Autonomous System Number (ASN) by authorities (i.e., RIPE NCC, APNIC or large ISPs ) and use the key pair to sign/validate the updates. In this way, the framework can verify the identities of AS participants that generated the updates. To guarantee the security of the transmitted updates, HTTPS connections can be built between two participants to prevent attackers accessing the updates \cite{zhang2020blockchain}. For meeting privacy requirements, many methods are proposed for the blockchain-based federated learning framework, such as \cite{weng2019deepchain} use threshold paillier algorithm, \cite{shayan2020biscotti} use Shamir’s secret sharing scheme, and \cite{chen2018machine} use differential privacy.

\textbf{Model storing:} Through the non-tamperable feature of blockchain, we can track/validate the uploading records and ensure the integrity of records. 
The initial block records the initial training model and task information. The consensus procedure is used to ensure that the global model update obtained by aggregation is unique, which can be implemented by Proof of Work (PoW) or Proof of Stake (PoS).
After the consensus procedure, the results of each global training epoch are stored as a transaction of the block in the blockchain. For instance, the $k$th global training results can be defined as $Y_{k} = [Hash(Y_{k-1}), T_{s}, T_{exp}, \Gamma, \mathbb{M}_{k}, SIG_{x}(Y_{k}), x]$, where $Hash(Y_{k-1})$ is the hash of previous block. Blocks in the blockchain are stored in a chain structure. The $T_s$ and $T_{exp}$ are the block generation time and expired time respectively. The $\Gamma$ is the global model update. The $SIG_{x}(Y_{k})$ is the block signature of the winner participant $x$ and $\mathbb{M}_{k}$ records the ASN of participants. Therefore, once a global epoch is finished, participants can download the global model update from the blockchain to update their local models. 

\begin{algorithm}[htbp]
  \caption{FL-RLD Training}
  \label{alg:federatedtraining}
  \begin{algorithmic}[1]
    \STATE \textbf{Input:} deployed ASes $M$, model $Q$, local training epoch $ce$, global training epoch $e$
    \STATE \textbf{Output:} global model update $\Gamma$
    \FOR{$k = 1$ to $ge$}
    \FOR{$\forall m \in M \textbf{ in Parallel}$}
    \STATE Obtaining local training data $D_m$ using algorithm \ref{alg:dataprocess} 
    \STATE local model $Q_m^{0} \gets Q$ 
    \FOR{$i = 1$ to $ce $}
    \STATE /* Training model */
    \STATE $Q_{m}^{i} \gets $ modelTraining$(Q_m^{i-1}, D_m)$ 
    \STATE /* Obtaining local model update */
    \STATE $\gamma_{m}^{i} \gets$ modelGetUpdate$(Q_m^{i-1},Q_m^{i})$
    
    \ENDFOR
    \STATE Uploading local model update $\gamma_{m}^{c}$
    \ENDFOR
    
    \FOR{$\forall m \in M \textbf{ in Parallel}$}
    \STATE /* Aggregating all local updates */
    \STATE $\Gamma_m^{k} \gets$ modelAggregate$(\{\gamma_{m}^{ce}| \forall m \in M\})$
    \ENDFOR
    \STATE /* Making consensus and obtaining global model update */ 
    \STATE $\Gamma \gets $ consensus$(\{\Gamma_m^{k}| \forall m \in M\})$
    \STATE Storing $\Gamma$ in the blockchain
    \ENDFOR
  \end{algorithmic}
\end{algorithm}

\subsection{Theoretical cost analysis}\label{subsec:cost}
Combining the training process in Fig.\ref{fig:fl-flow} and training algorithm  shown in Algorithm \ref{alg:federatedtraining},
the total cost of a training task can be divided into three parts: local computation cost (step \ding{172}, \ding{173}, \ding{175}), global communication cost (step \ding{174}, \ding{176}) and storing cost (step \ding{178}). 

\textbf{Local computation cost:} 
  Let $L^{i}(D_{m})$ be the local training cost of the participant $m \in M$ in $i$th local training epoch where $D_{m}$ is the local training data of $m$ and $a_{k}$ be the model aggregation cost in $k$th global training.
  So, the local computation cost is $\sum_{i}^{ce} L^{i}(D_{m}) + a_{k}$ where $ce$ is the local training epochs, $D_m$ is the local dataset of $m$. 

\textbf{Global communication cost:} The global communication cost includes two parts, exchanging local updates and making consensus about aggregated global updates. First, let $\delta_{m}^{k}$ be the local update of $m$ in the $k$th global training epoch. Assuming the communication cost between any two participants for $m$ is $\Delta(\delta_{m}^k)$. Thus, the cost of broadcasting the local model update for $m$ is defined as $ (|M|-1) \Delta (\delta_{m}^k) $ where $|M|$ is the number of participants. Then, let $\zeta^{k}(|M|)$ be the consensus cost of $k$th global epoch. Finally, the total global communication cost is $\sum_{k}^{ge} \sum_{m \in M} ( (|M|-1) \Delta (\delta_{m}^k) + \zeta^k(|M|))$ where $ge$ is the global training epochs.

\textbf{Storing cost:} The global updates in each global training epoch are stored in blocks. The storing cost is related to the size of model updates, so the storing cost is $\sum_{k}^{ge} f(\Gamma^{k})$ where the $\Gamma^{k}$ is the global model update in $k$th global training epoch.

Therefore, the total cost of a training task can be presented as following:
\begin{equation}
  \begin{split}
    total\_cost &= \sum_{k}^{ge} \sum_{m \in M} ( \overbrace{(\sum_{i}^{ce} L^{i}(D_m)) + a_{k}}^{\text{Local computation cost}} \\ 
    &+ \underbrace{\overbrace{(|M|-1)\Delta(\delta_m^k)}^{\text{Exchanging update cost}}+\overbrace{\zeta^k(|M|)}^{\text{Consensus cost}}}_{\text{Global communication cost}} \\
    &+ \underbrace{f(\Gamma^k)}_{\text{Storing cost}} )
  \end{split}
  \label{eq:cost}
\end{equation}
The parameters used in the Equation (\ref{eq:cost}) are summarized in Table \ref{tab:notations}.
\begin{table}[htbp]
  \caption{List of notations used in Equation (\ref{eq:cost})}
  \label{tab:notations}
  \begin{tabular}{cl}
  \hline
  \rowcolor[HTML]{C0C0C0} Parameter                & Meaning                                                                         \\ \hline
  $ge$                     & Global training epochs                                                          \\ 
  $ce$                     & Local training epochs                                                           \\ 
  $M, |M|$                 & Deployed AS set, the number of $M$                                              \\ 
  $D_m$                    & Training data of deployed AS $m$                                                \\ 
  $L^{i}(D_m)$             & Training cost of $m$ in $i$th local training epoch                              \\ 
  $a_k$                    & Aggregation cost in $k$th global training epoch                                 \\ 
  $\delta_{m}^{k}$         & Local model update of $m$ in $k$th global training epoch                        \\ 
  $\Delta(\delta_{m}^{k})$ & Cost of broadcasting $\delta_{m}^{k}$ between any two participants \\ 
  $\zeta^{k}(|M|)$         & Consensus cost                                                                  \\ 
  $\Gamma^{k}$             & Global model update in $k$ the global training epoch                            \\ 
  $f(\Gamma^{k})$          & Cost of storing global model update in the blockchain                           \\ \hline
  \end{tabular}
  \end{table}
\subsection{Factors affecting deployment}\label{subsec:factors}
Here, to promote the deployment, we discuss which factors can affect the detection effectiveness of the proposed method.
The Internet is modeled as a graph $G(V,E)$ where $V$ is a set of all ASes in the Internet and $E$ represents the direct links between ASes. In the graph, $M$ is a set of ASes that have deployed the proposed detection system, where $M \subset V$. For each AS $v \in M$, the local training data of $m$ is represented as $D_v$ where $|D_v| \geq 0 $. Thus, the shared local training data of $M$ can be defined as $\mathbb{D} = \{D_v | \forall v \in M \}$ and is used to federally train a global model to identify the input AS triple is malicious or regular. Thus, the aim of the global model is to memorize the mappings of AS triples and their label of local training data as much as possible. The accuracy that the global model's can correctly identify a malicious AS triple in $\mathbb{D}$ is set as $\theta$, where $0 \leq \theta \leq 1$.

\begin{theorem}
    If $\theta = 1$, consider AS $v_{k+1} \in M$ receives a leaking announcement $\mathcal{R}=\{v_1, ..., v_{h}, ..., v_{k}\}$ and the leaking AS is $v_{h}$. If $v_h \in M$, then the AS triple $(v_{h-1}, v_{h}, v_{h+1})$ and its label are in $\mathbb{D}$. Therefore, AS $v_{k+1}$ can identify that $\mathcal{R}$ is a malicious announcement. For $v_h \notin M$, if $v_{h+1} \in M$ and AS triple $(v_{h-1}, v_{h}, v_{h+1})$ in $\mathbb{D}$, then AS $v_{k+1}$ can identify that $\mathcal{R}$ is malicious.
\end{theorem}

When $\theta = 1$, the global model can identify every AS triple in $\mathbb{D}$ is malicious or not. So, once AS $v \in M$ receives an announcement with a malicious AS triple in $\mathbb{D}$, it will correctly detect the malicious announcement.
If $\theta < 1$, then the number of malicious AS triples that the global model can identify is $\theta \cdot |\mathbb{D}|$. 

Therefore, it can be concluded that the performance of route leak detection depends on two parts, accuracy of the global model and the number of malicious AS triples. Under the same accuracy of the global model, more route leaks will be detected as the number of malicious AS triples in $\mathbb{D}$ increases.

\begin{table*}[!h]
    \centering
    \caption{The triple distribution of different groups}
    \label{tab:groups}
    \scalebox{1}{
        \begin{tabular}{lccccc}
        \hline
         & Data size & Anomaly & Regular & Anomaly \% & Regular \% \\ \hline
        \rowcolor[HTML]{C0C0C0} 
        Group 1 (unbalanced data size + unbalanced class distribution) & 13550     & 12224   & 1326    & 90.21\%    & 9.79\%     \\
        Client1 (51.19\%)                                              & 6936      & 6192    & 744     & 89.27\%    & 10.73\%    \\
        Client2 (30.92\%)                                              & 4189      & 3913    & 276     & 93.41\%    & 6.59\%     \\
        Client3 (0.51\%)                                               & 69        & 33      & 36      & 47.83\%    & 52.17\%    \\
        Client4 (14.18\%)                                              & 1922      & 1680    & 242     & 87.41\%    & 12.59\%    \\
        Client5 (3.20\%)                                               & 434       & 406     & 28      & 93.55\%    & 6.45\%     \\
        \rowcolor[HTML]{C0C0C0} 
        Group 2 (balanced data size + unbalanced class distribution)   & 63468     & 51066   & 12402   & 80.46\%    & 19.54\%    \\
        Client1 (19.77\%)                                              & 12549     & 12099   & 450     & 96.41\%    & 3.59\%     \\
        Client2 (20.69\%)                                              & 13134     & 12158   & 976     & 92.57\%    & 7.43\%     \\
        Client3 (19.25\%)                                              & 12218     & 7606    & 4612    & 62.25\%    & 37.75\%    \\
        Client4 (19.49\%)                                              & 12369     & 10205   & 2164    & 82.51\%    & 17.50\%    \\
        Client5 (20.79\%)                                              & 13198     & 8998    & 4200    & 68.18\%    & 31.82\%    \\
        \rowcolor[HTML]{C0C0C0} 
        Group 3 (unbalanced data size + balanced class distribution)   & 416348    & 208174  & 208174  & 50.00\%    & 50.00\%    \\
        Client1 (8.58\%)                                               & 35712     & 17856   & 17856   & 50.00\%    & 50.00\%    \\
        Client2 (35.93\%)                                              & 149580    & 74790   & 74790   & 50.00\%    & 50.00\%    \\
        Client3 (43.45\%)                                              & 180904    & 90452   & 90452   & 50.00\%    & 50.00\%    \\
        Client4 (10.40\%)                                              & 43316     & 21658   & 21658   & 50.00\%    & 50.00\%    \\
        Client5 (1.64\%)                                               & 6836      & 3418    & 3418    & 50.00\%    & 50.00\%    \\
        \rowcolor[HTML]{C0C0C0} 
        Group 4 (balanced data size + balanced class distribution)     & 17090     & 8512    & 8578    & 49.81\%    & 50.19\%    \\
        Client1 (20\%)                                                 & 3418      & 1761    & 1657    & 51.52\%    & 48.48\%    \\
        Client2 (20\%)                                                 & 3418      & 1672    & 1746    & 48.92\%    & 51.08\%    \\
        Client3 (20\%)                                                 & 3418      & 1724    & 1694    & 50.44\%    & 49.56\%    \\
        Client4 (20\%)                                                 & 3418      & 1679    & 1739    & 49.12\%    & 50.88\%    \\
        Client5 (20\%)                                                 & 3418      & 1676    & 1742    & 49.04\%    & 50.97\%    \\ \hline
        \end{tabular}
    }
    \end{table*}

\section{Experiments and analysis}\label{sec:experiments}
In this section, we give a description about experiment setup and show the experiment results. 
To learn more about the possible route leaks, we first explore the features regarding generated triples, such as the proportion of malicious triples and regular triples, and how the two types of triples have changed over the past four years. Then, experiments are conducted to evaluate the performance of the proposed detection method. 

\subsection{Experiment setup}\label{sec:experimentsetup}

\textbf{Topology:}
The BGP topology data used in the evaluation is collected from the CAIDA January 2021 AS relationship dataset \cite{caidaRelation} of IPv6, which has 12,721 ASes and 173,462 AS links.
We use the method introduced in Section \ref{subsec:datageneration} to generate triples and their labels for ASes in the network. 


\textbf{Implementation details:} The proposed method is implemented by Python and Keras \cite{ketkar2017introduction}, and the initial training model is a simple LSTM network. It includes a LSTM layer with input size (1, 96) and output size (1, 128), a hidden layer with output size (1, 64) and ReLU \cite{nair2010rectified} activation function, and an output layer with output size (1, 2) and Softmax activation function. For each 2-bit vector of the output layer, if the first bit is larger than the second one, then it is predicted as a regular triple. Otherwise it is predicted as a malicious triple. The model uses Adam optimizer \cite{kingma2014adam} as the optimization process and the batch size is 32. The learning rate of the model is set as 0.001 and the FedAvg algorithm \cite{mcmahan2017communication} is used to aggregate local updates. In the training data, each ASN in the generated triples is embedded as a 32-bit vector by converting decimal to binary. The local training epoch is set as 2 and the global training epoch is around 70 to 100.

\textbf{Training data details:} In our experiments, we consider two aspects of local training data that may influence the results, balanced/unbalanced data size and balanced/unbalanced class distribution. We select 4 groups of federated learning participants to test. Each group has 5 participants. We represent each participant of a group with Client1, Client2, Client3, Client4, and Client5. The details about the groups are illustrated in Table \ref{tab:groups}. The data size refers to the number of triples of the participant. For example, in group 1, participants have different sizes of local training data, and the number of malicious and regular triples are not equal. In group 3, participants have an equal number of malicious and regular triples but their total number of triples are different.
We run tests on the above four groups. For a clear description, we use $\mathcal{D}_{g,c}$ to denote the local training data of participant $c$ in the group $g$. For instance, the local training data of Client1 in Group 1 is $\mathcal{D}_{1,1}$. In each experiment, all local training data from a group is used to test the trained model. 

\textbf{Evaluation metric:} Four standard metrics, \textit{Accuracy, Precision, Recall, F1score} are used as evaluation metrics for detection performance. \textit{Accuracy} shows the ratio of correctly predicted triples to the total triples. \textit{Precision} and \textit{Recall} display the ratio of correctly predicted malicious triples and regular triples respectively. \textit{F1score} is the average of \textit{Precision} and \textit{Recall}. 
Their definitions are as follows. 
\begin{equation}
    Accuracy = \frac{TP + TN}{TP + FP + TN + FN}
\end{equation}
\begin{equation}
    Precision = \frac{TP}{TP + FP}
\end{equation}
\begin{equation}
    Recall = \frac{TP}{TP + FN}
\end{equation}
\begin{equation}
    F1score = 2 \frac{Precision * Recall}{Precision + Recall}
\end{equation}
where True Positives (TP) and False Positives (FP) are the number of true malicious triples that the model predicts as anomaly and regular respectively. The True Negatives (TN) and False Positives (FP) are the number of true regular triples that the model predicts as regular and anomaly respectively.

\textbf{Comparative methods:} We use FL-RLD to represent the proposed method, and \textit{CL} to represent the central learning method. The difference between CL and FL-RLD is that CL transports all local training data of participants of a group to a single server for model training while FL-RLD keeps the data training local. The \textit{C1, C2, C3, C4}, and \textit{C5} represent the single AS learning method with participant Client1, Client2, Client3, Client4, and Client5 respectively. So, the single learning method can only utilize the local training data of a single AS to train the model. For example, the training data of C1 in group 1 uses the local training data of corresponding AS 
is $\mathcal{D}_{1,1}$ and the training data of CL is $\{\mathcal{D}_{1,i}| i =1,2,...,5\}$. The training model of FL-RLD, CL and C1/C2/C3/C4 are all the same. 

Traditional detection methods work by storing AS business relationships in various ways and filtering out routes that aren't matched (i.e., building a RPKI-like repository to store the routing customer-provider authority objects \cite{azimovverification}, marking the "Down-only" routes \cite{sriram2017methods} to prevent forwarding routes to upstream providers or peers).
Hence, we modeled three different methods \textit{ML-random, ML-0, ML-1} based on the above analysis to compare with FL-RLD. In these three methods, they all build a global repository that all participants of a group directly share their AS relationships. Ideally, if all relationship information of the AS triple in test data is in the global repository, output the correct result. Otherwise, they respond differently: 1) ML-random will randomly output a result. 2) ML- 0 will mark this AS triple as malicious. 3) ML-1 will mark this AS triple as regular.

\subsection{Performance}

\begin{figure*}[!h]
    \begin{center}
    \begin{subfigure}{0.24\textwidth}
        \centerline{\includegraphics[scale=0.25]{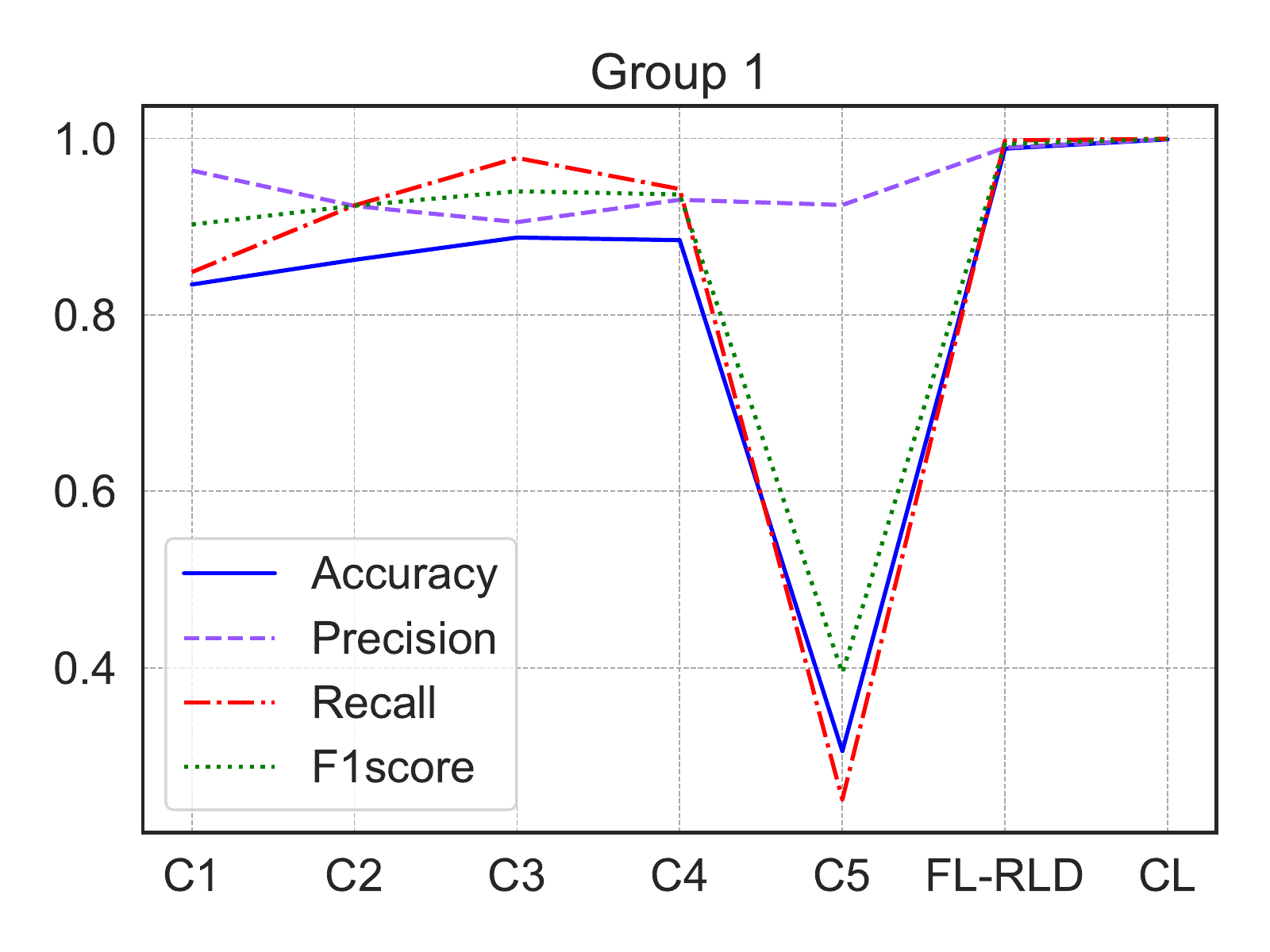}}
        \caption{Unbalanced data size + unbalanced class distribution}
        \label{fig:sgroup1}
    \end{subfigure}
    \begin{subfigure}{0.24\textwidth}
        \centerline{\includegraphics[scale=0.25]{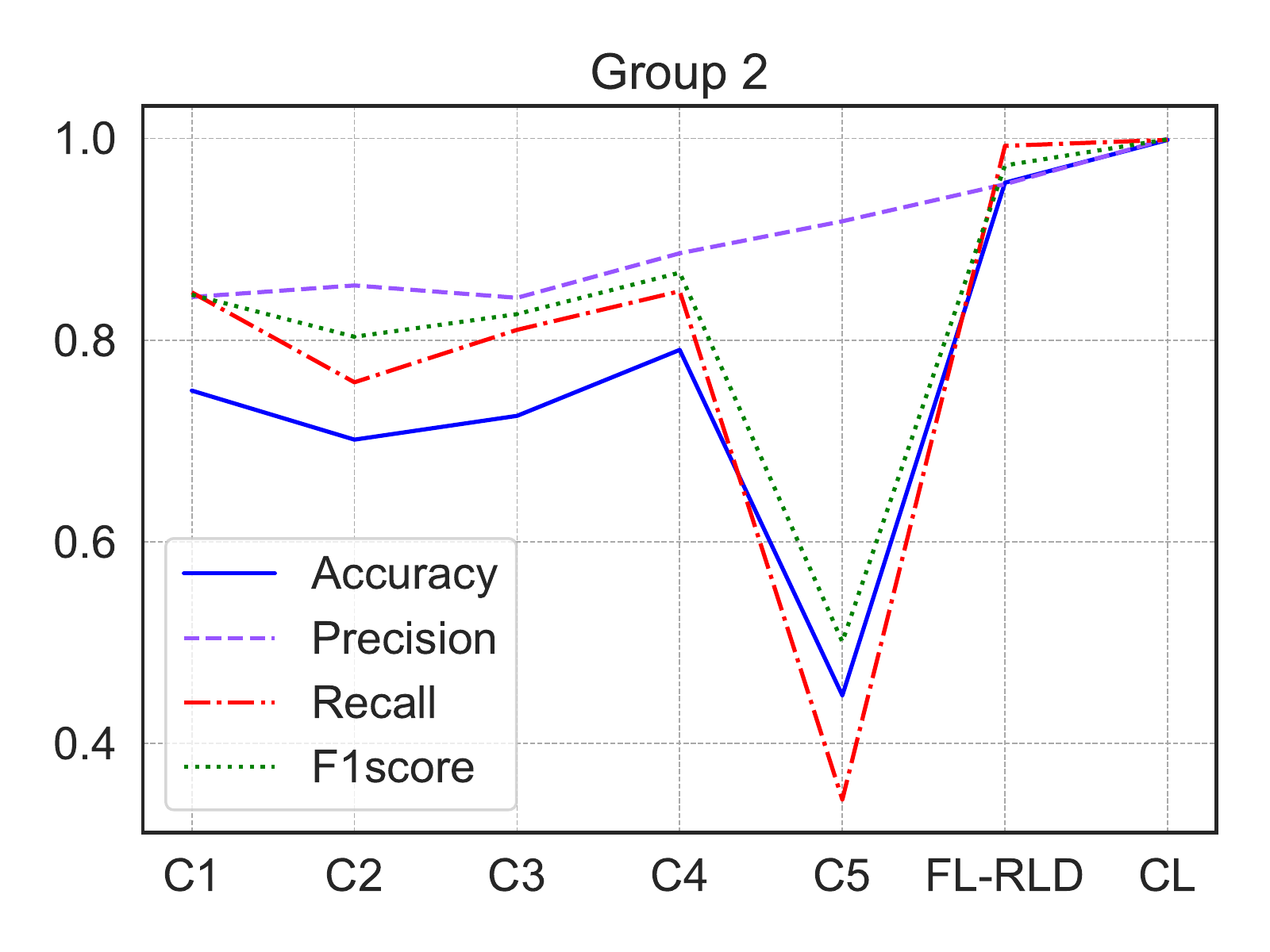}}
        \caption{Data size balance + unbalanced class distribution}
        \label{fig:sgroup2}
    \end{subfigure}
    \begin{subfigure}{0.24\textwidth}
        \centerline{\includegraphics[scale=0.25]{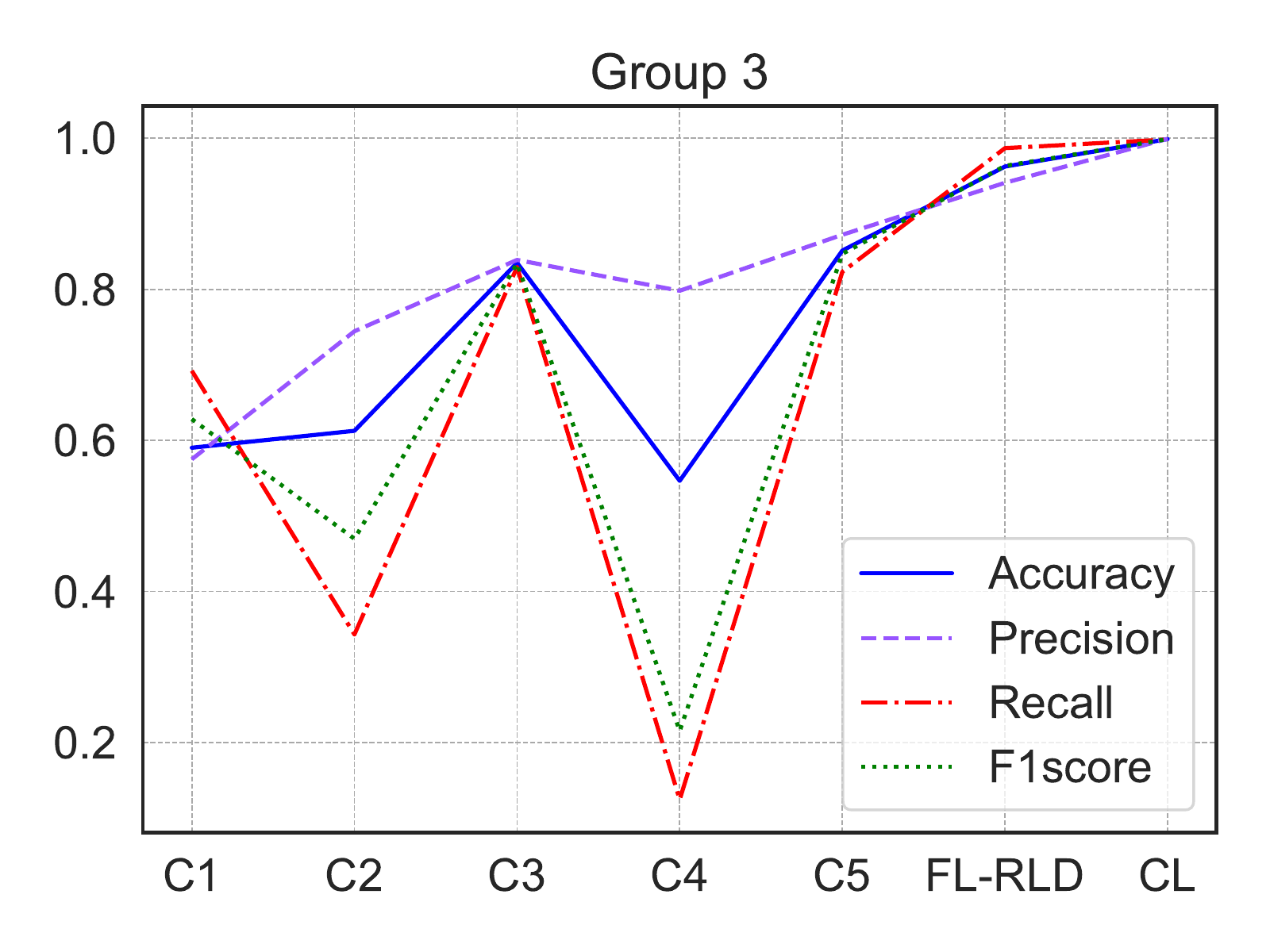}}
        \caption{Unbalanced data size + balanced class distribution}
        \label{fig:sgroup3}
    \end{subfigure}
    \begin{subfigure}{0.24\textwidth}
        \centerline{\includegraphics[scale=0.25]{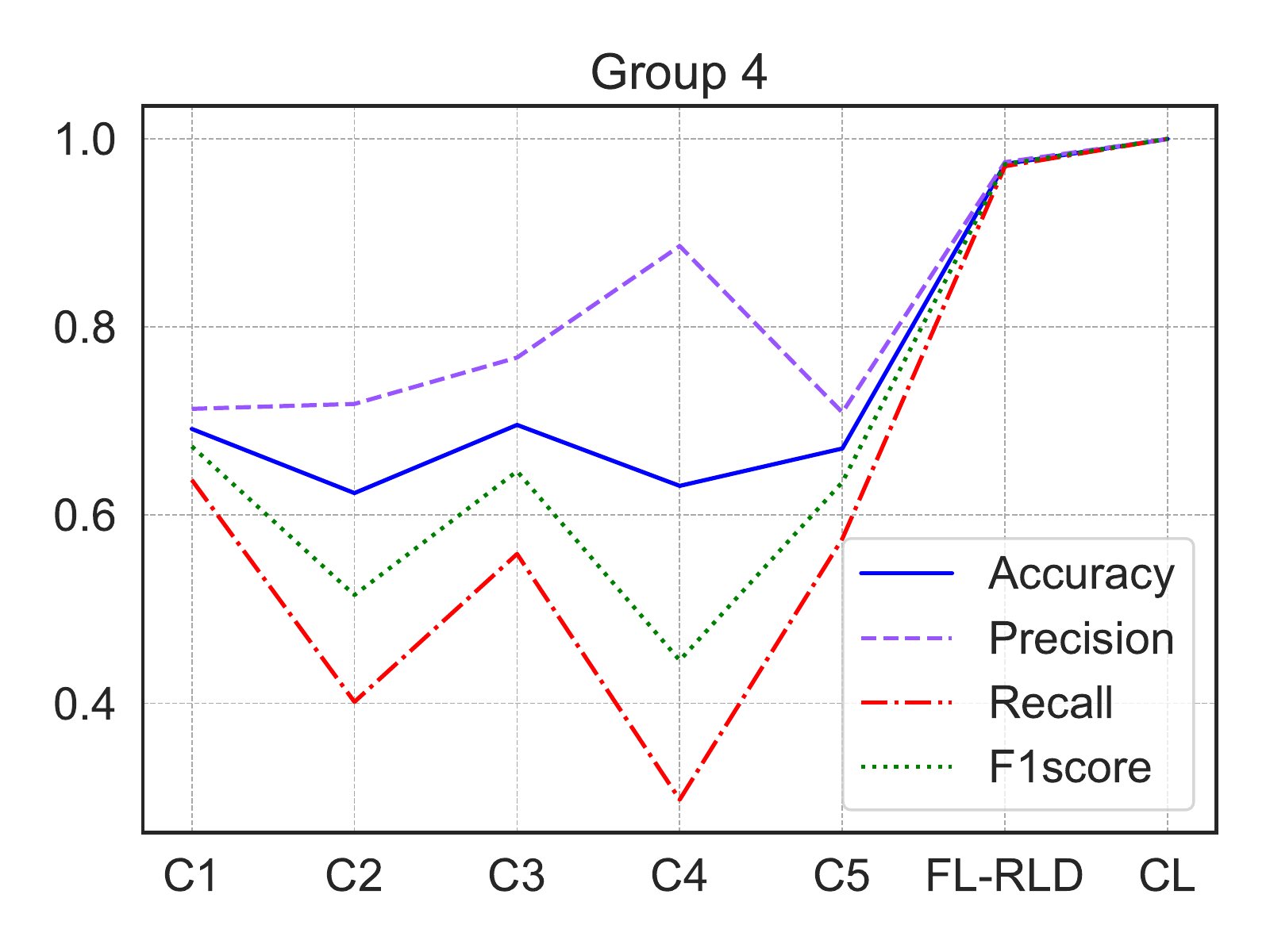}}
        \caption{Data size balance + balanced class distribution}
        \label{fig:sgroup4}
    \end{subfigure}
    \caption{The Performance of FL-RLD method compared with single AS learning method (C1, C2, C3, C4, C5) and Central Learning (CL) method}
    \label{fig:metrics}
    \end{center}
\end{figure*}

\begin{figure*}[!h]
    \begin{center}
    \begin{subfigure}{0.24\textwidth}
        \centerline{\includegraphics[scale=0.25]{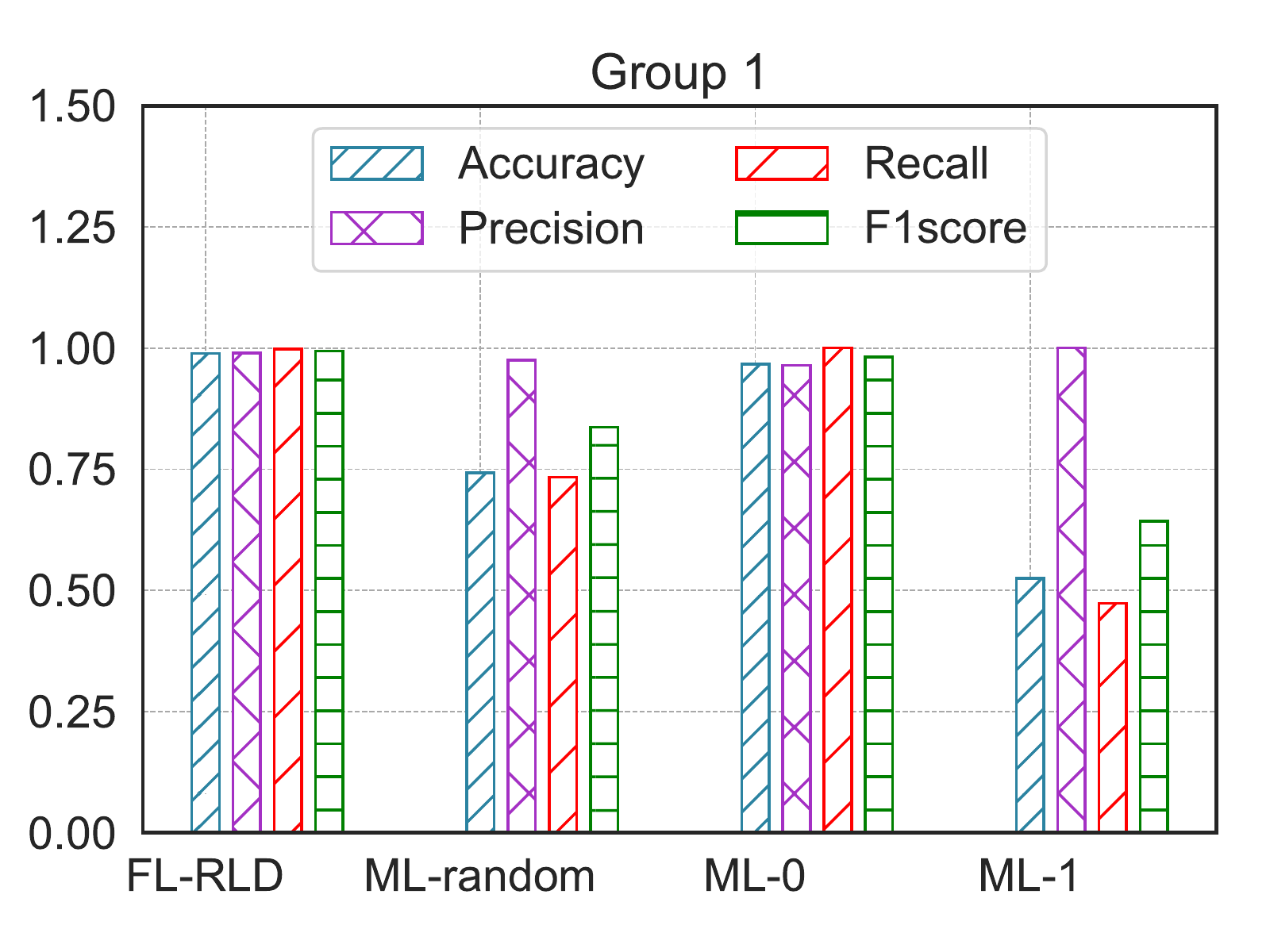}}
        \caption{Anomaly 90.21\% vs Regular 9.79\%}
        \label{fig:group11}
    \end{subfigure}
    \begin{subfigure}{0.24\textwidth}
        \centerline{\includegraphics[scale=0.25]{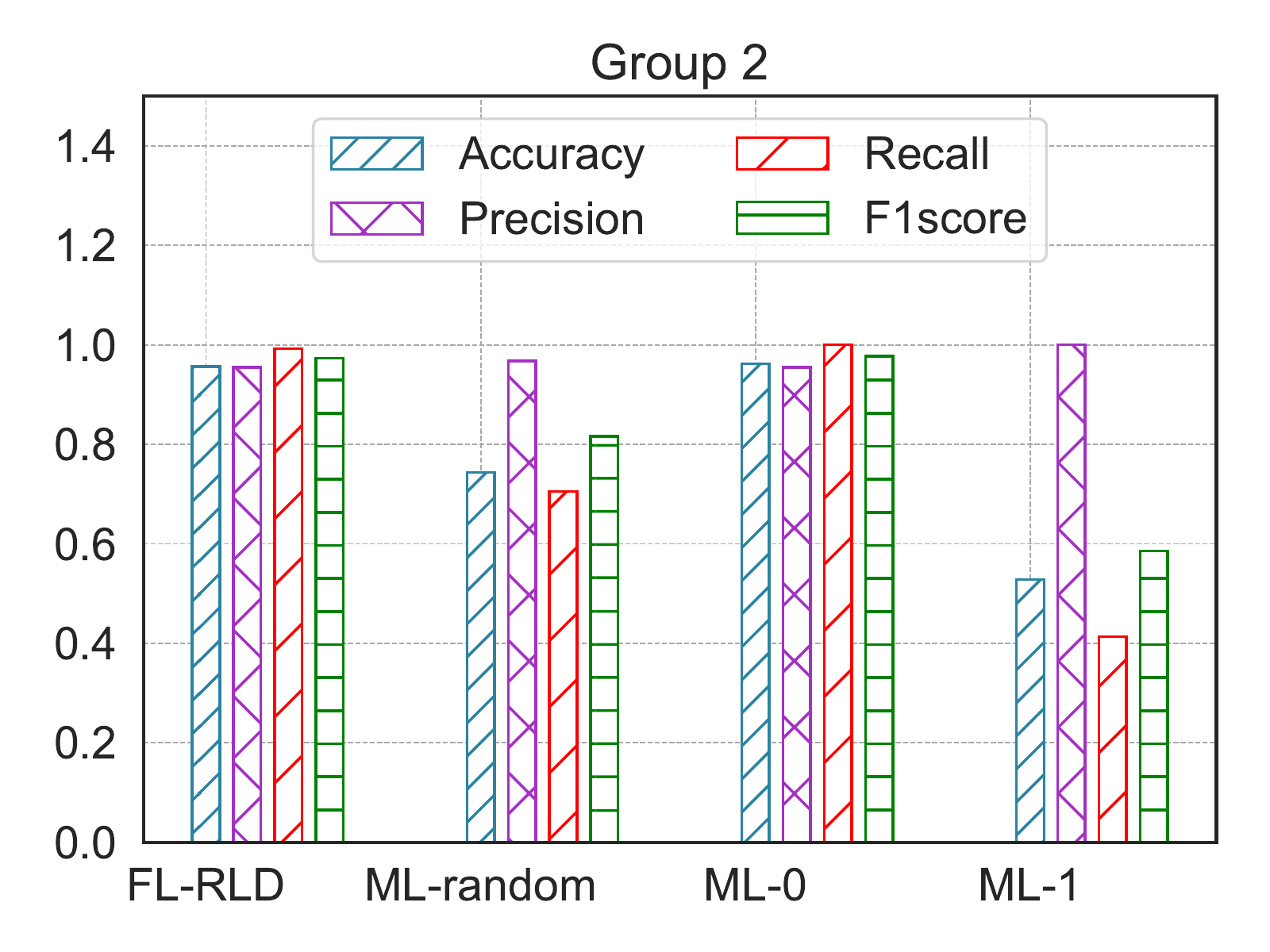}}
        \caption{Anomaly 80.46\% vs Regular 19.54\%}
        \label{fig:group12}
    \end{subfigure}
    \begin{subfigure}{0.24\textwidth}
        \centerline{\includegraphics[scale=0.25]{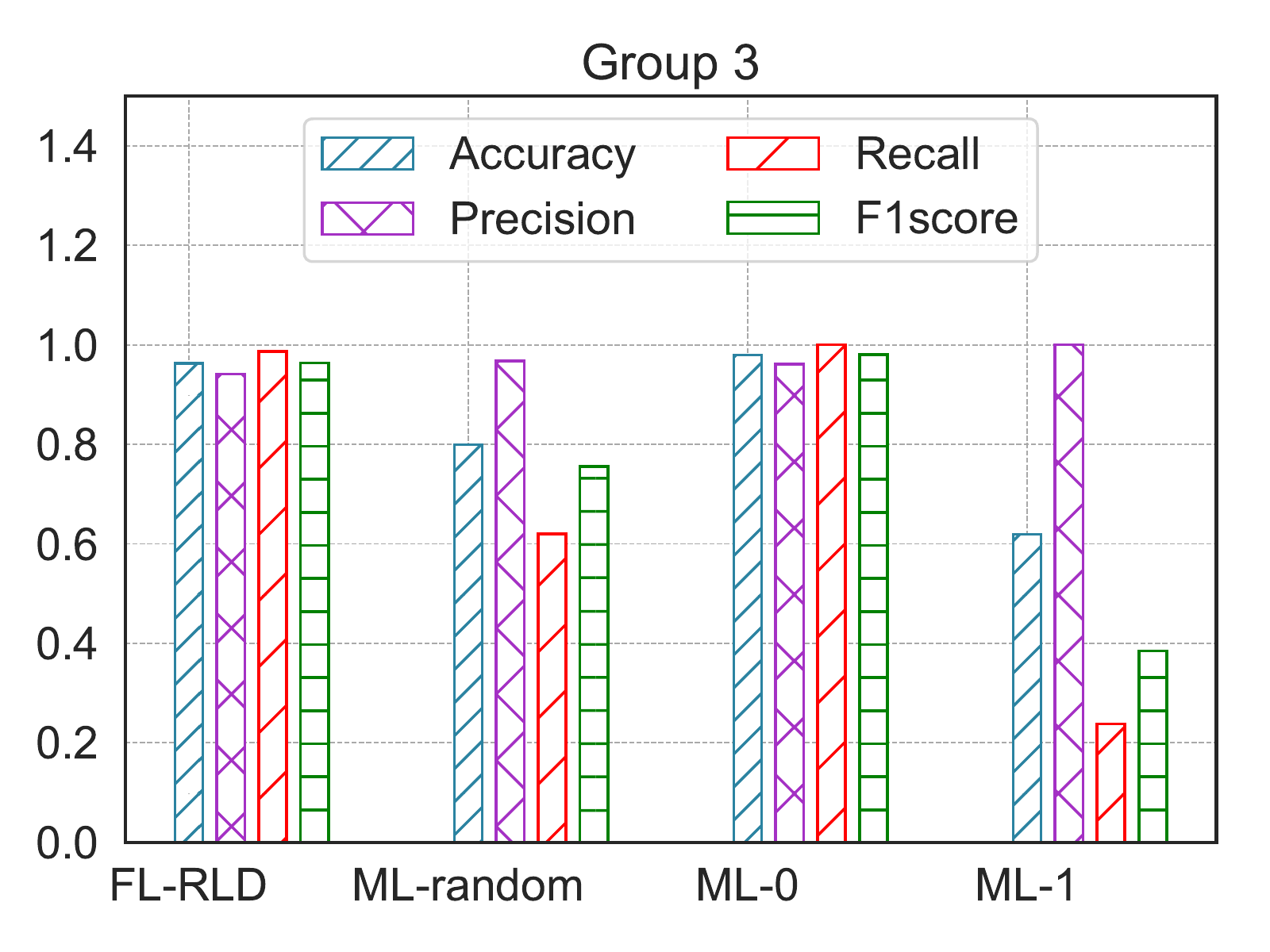}}
        \caption{Anomaly 50\% vs Regular 50\%}
        \label{fig:group13}
    \end{subfigure}
    \begin{subfigure}{0.24\textwidth}
        \centerline{\includegraphics[scale=0.25]{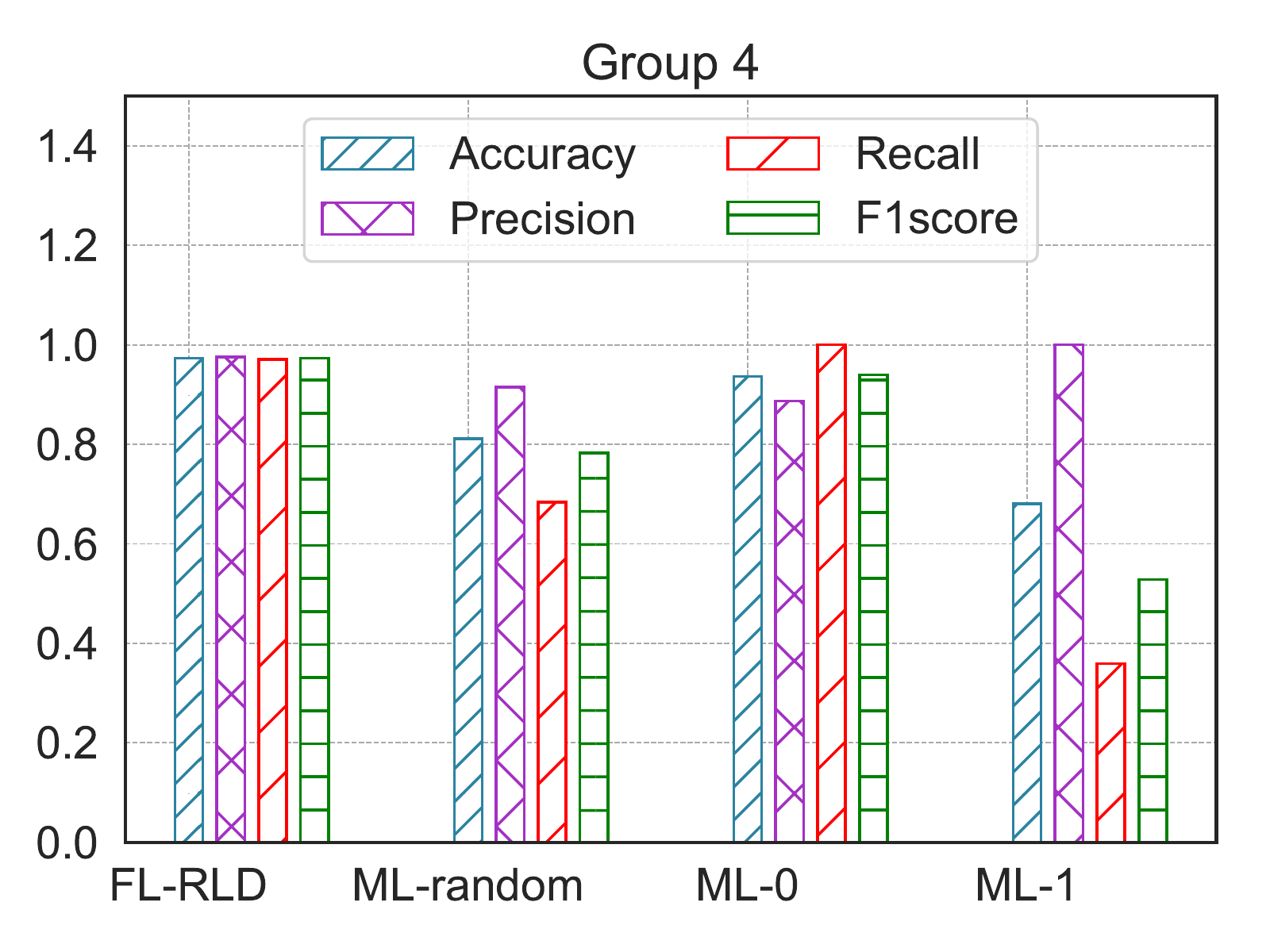}}
        \caption{Anomaly 49.81\% vs Regular 50.19\%}
        \label{fig:group14}
    \end{subfigure}
    \caption{The performance comparison of FL-RLD and other methods.}
    \label{fig:groups}
    \end{center}
\end{figure*}

\textbf{Multiple ASes \textit{vs.} Single AS:}  First, a comparison of a global model trained by multiple ASes and a model trained by a single AS is carried out. The results are shown in Fig.\ref{fig:metrics}. 
The Fig.\ref{fig:sgroup1} and Fig.\ref{fig:sgroup2} are results under datasets with unbalanced class distribution, while Fig.\ref{fig:sgroup3} and Fig.\ref{fig:sgroup4} are results under datasets with balanced class distribution. 
In Fig.\ref{fig:metrics}, FL-RLD performs better than C1, C2, C3, C4 and C5 in all evaluation metrics under different groups of datasets, which provides an incentive for ASes to join federated learning. For example, in Fig.\ref{fig:sgroup2}, the \textit{Accuracy} of C1 to C5 are all lower than 0.8 but when they join the federated learning, the \textit{Accuracy} is more than 0.95.
The results also show that the difference of FL-RLD and CL in the performance is small (e.g., less than 0.06 \textit{Accuracy} in group 1). 
\begin{figure}[htbp]
    \begin{center}
    \begin{subfigure}{0.45\textwidth}
        \centerline{\includegraphics[scale=0.45]{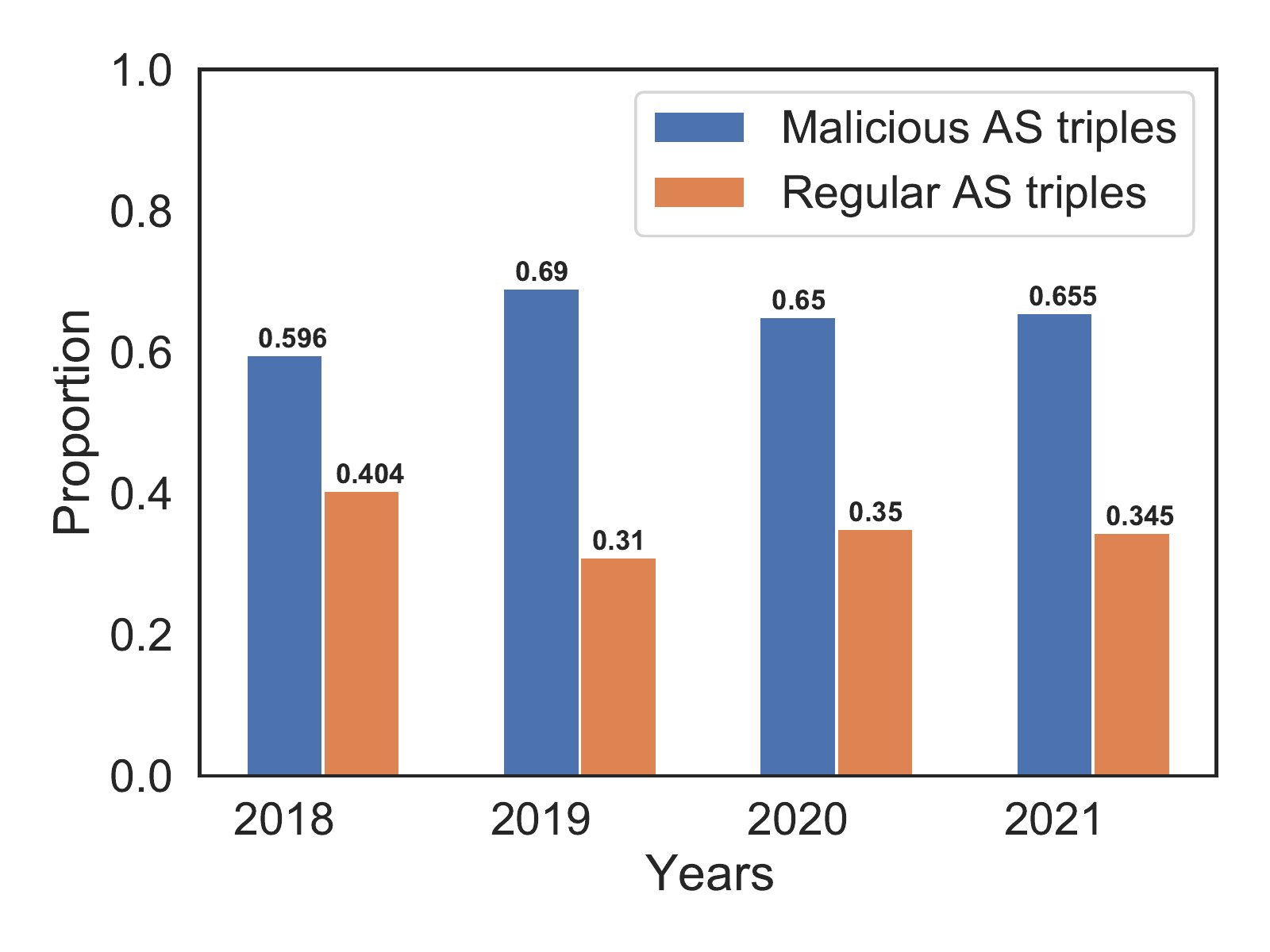}}
        \caption{Distribution of proportion of malicious and regular generated triples of local training data under different years. \textit{Please note the results are distribution of possible triples and are not distribution of triples appeared in the actual routing announcements.} }
        \label{fig:yeardis}
    \end{subfigure}
    \begin{subfigure}{0.45\textwidth}
        \centerline{\includegraphics[scale=0.45]{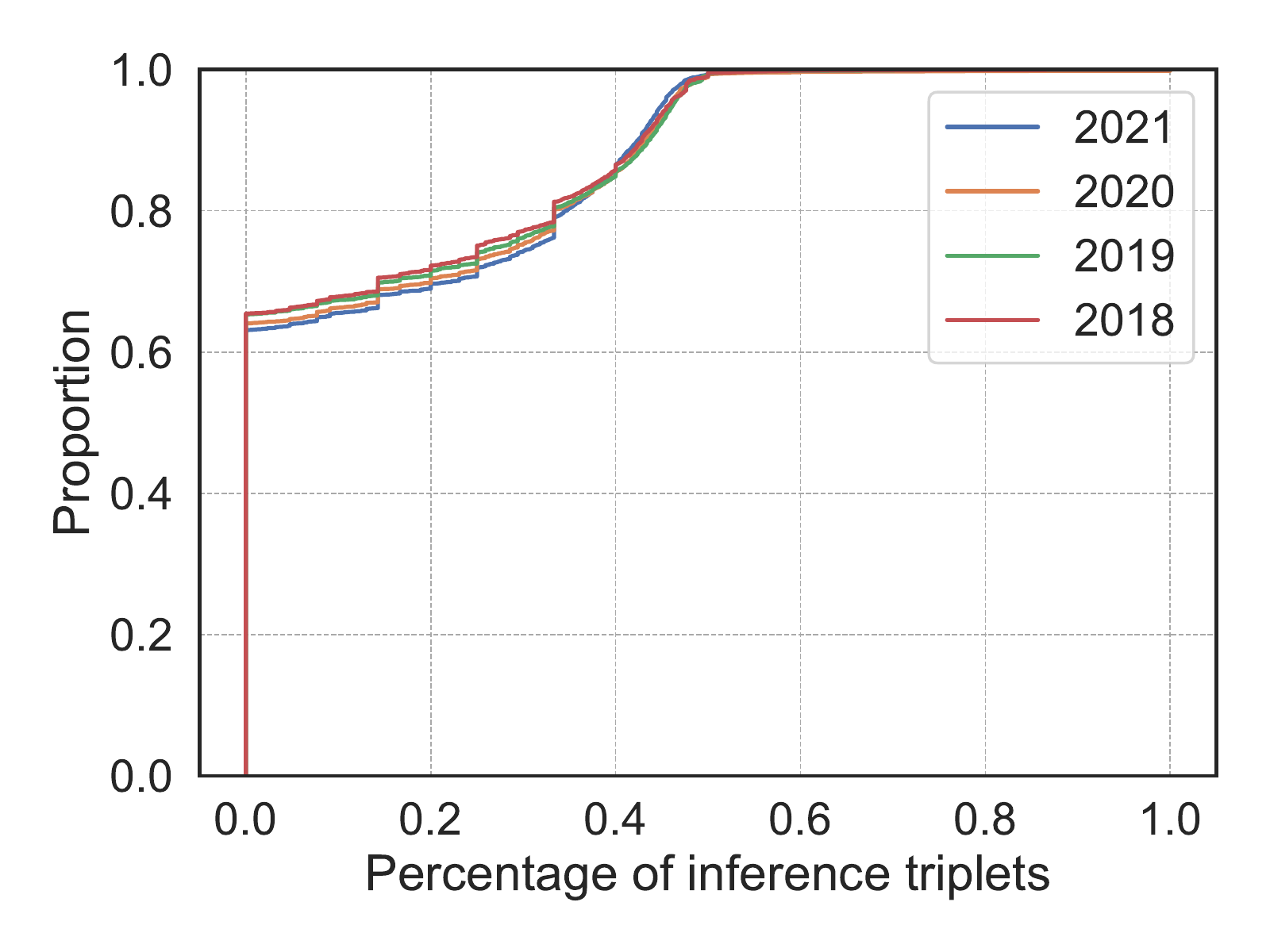}}
        \caption{CDF of the percentage of inference triples to total triples in different years.}
        \label{fig:inference}
    \end{subfigure}
    \caption{Distribution of triples in different years.}
    \label{fig:years}
    \end{center}
\end{figure}

\textbf{Global repository \textit{vs.} FL-RLD:} In Fig.\ref{fig:groups}, we compare FL-RLD with ML-random, ML-0 and ML-1, where these three comparative methods are based on sharing a global repository of business relationships. As introduced previously, the main difference of these three methods is that they respond differently when the AS relationship information is not in the training data. 
The results in Fig.\ref{fig:groups} show that the performance of FL-RLD and ML-0 is better than others and FL-RLD performs better on average than ML-0. 
The ML-1 performs the worst because the number of triples marked as malicious are higher than that of regular triples, while the ML-1 classifies all unknown triples as regular, thus making the Recall low. The ML-0 classifies all unknown triples as malicious, so its performance is better than ML-1. However, the Precision of ML-1 is low, which makes high false alarms. Thus, the FL-RLD that performs well on both Precision and Recall is more recommended.

\begin{figure}
    \centerline{\includegraphics[scale=0.5]{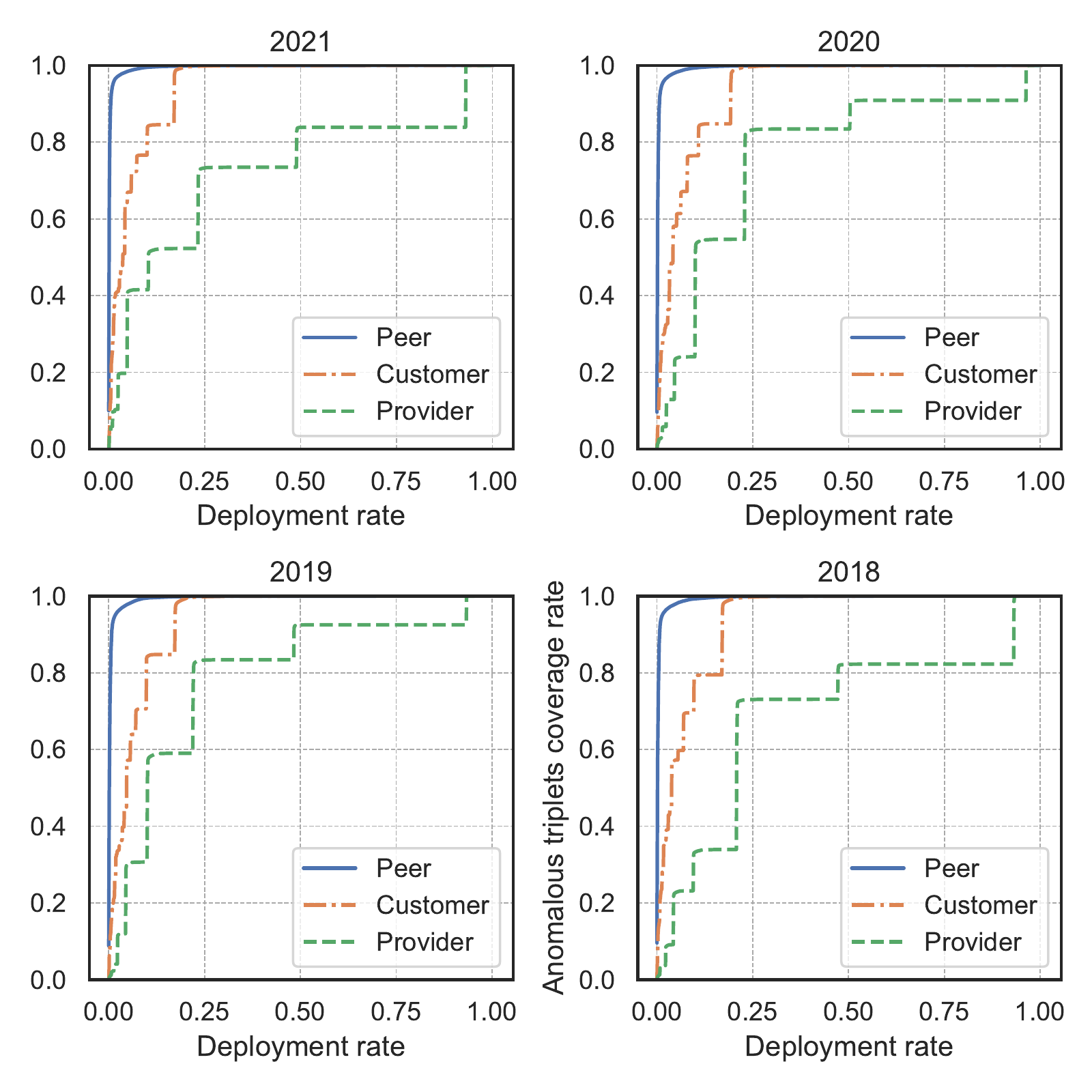}}
    \caption{The proportion of malicious triples in the training data to total malicious triples in the network using different deployment strategies (Peer, Customer, and Provider). }
    \label{fig:yearsDeploy}
\end{figure}
\subsection{Deployment analysis}
As analyzed in Section \ref{subsec:factors}, the higher number of malicious AS triples are contained in the training data, the more route leaks can be detected by FL-RLD. Thus, to facilitate the deployment, we first analyze the distribution of AS triples over time and study the relationship between the number of deployed ASes and the number of route leaks that can be detected under different deployment strategies if the accuracy $\theta$ of the global model is 1.

\textbf{Malicious AS triples \textit{vs.} Regular AS triples:} Except for the 2021 topology used above, we also collect other three topologies, 2020, 2019, 2018, to study the distribution of AS triples over time. They are also from CAIDA AS relationship dataset of IPv6. 
Fig.\ref{fig:years} shows the proportion of malicious and regular triples of the local training data under different topology data grouped by year. As we can see, the results show that the number of malicious triples (around 60\%-70\%) are more than regular triples (around 30\%-40\%) in the local training data. This is because the number of peer-to-peer relationships are more than provider-to-customer relationships, which makes extensive possible route leaks.
Please note that even the number of malicious AS triples are higher than regular AS triples, but actually most of ASes in the network act normally and only a small number of ASes act maliciously. Therefore, there are not so many malicious AS triples in the actual routing announcements.

\textbf{Direct AS triples \textit{vs.} Inference AS triples:} Fig.\ref{fig:inference} depicts CDFs of the percentage of inference triples to total triples. In the about 80\% ASes, around 30\% triples are inference triples where neighbors are the possible leakers. The data of the past four years from 2018 to 2021 show a similar result. It indicates that our generation method can enrich training data well by extending properly proportioned AS triples based on the known routing policies.

\textbf{Deployment strategies:} Fig.\ref{fig:yearsDeploy} shows the distribution of malicious triples in the training data to total malicious triples under different deployment strategies, where Peer/Customer/Provider selects ASes with the largest number of peers/customers/providers to deploy first. The results show that Peer deployment strategy can cover the most number of malicious triples than other two strategies with the same deployment rate. For example, in the data of 2021, if using Peer deployment strategy, it only needs to deploy 875 ASes (0.06878 deployment rate) to cover 99\% total malicious triples while Customer deployment strategy can only reach 72\% coverage rate with the same deployment rate. It also indicates that ASes with a large number of peers have a large number of possible malicious triples.


\section{Conclusion}\label{sec:conclusion}
This paper studied the problem of route leak detection in the inter-domain network and proposed a privacy-preserving method using blockchain-based federated learning to collaboratively train models with accurate routing policy information for route leak detection. Compared with traditional route leak detection methods, the proposed method considers the privacy of AS business relationships. To solve the lack of a ground truth problem in route leaks, it provides a self-validation scheme by labeling AS triples as malicious or regular using local routing policies. The evaluation results show that the proposed method can improve the performance of a single AS in detecting route leaks, and there is slight differences in results (e.g., around 0.06 in \textit{Accuracy}) between the federated learning method and the central learning method, in which all local training data are gathered and trained together. In the analysis of FL-RLD deployment, it is found that AS with more peers are more likely to have more possible route leaks and can contribute more on route leak detection if they join the federated learning.

\section*{Acknowledgement}
This work was supported by the National Key R\&D Program of China (No. 2018YFB1800404).

\ifCLASSOPTIONcompsoc


\ifCLASSOPTIONcaptionsoff
  \newpage
\fi



%

\bibliographystyle{IEEEtran} 
\bibliography{refs.bib}

\begin{thebibliography}{10}
\providecommand{\url}[1]{#1}
\csname url@samestyle\endcsname
\providecommand{\newblock}{\relax}
\providecommand{\bibinfo}[2]{#2}
\providecommand{\BIBentrySTDinterwordspacing}{\spaceskip=0pt\relax}
\providecommand{\BIBentryALTinterwordstretchfactor}{4}
\providecommand{\BIBentryALTinterwordspacing}{\spaceskip=\fontdimen2\font plus
\BIBentryALTinterwordstretchfactor\fontdimen3\font minus
  \fontdimen4\font\relax}
\providecommand{\BIBforeignlanguage}[2]{{%
\expandafter\ifx\csname l@#1\endcsname\relax
\typeout{** WARNING: IEEEtran.bst: No hyphenation pattern has been}%
\typeout{** loaded for the language `#1'. Using the pattern for}%
\typeout{** the default language instead.}%
\else
\language=\csname l@#1\endcsname
\fi
#2}}
\providecommand{\BIBdecl}{\relax}
\BIBdecl

\bibitem{luckie2013relationships}
M.~Luckie, B.~Huffaker, A.~Dhamdhere, V.~Giotsas, and K.~Claffy, ``As
  relationships, customer cones, and validation,'' in \emph{Proceedings of the
  2013 conference on Internet measurement conference}, 2013, pp. 243--256.

\bibitem{gao2001inferring}
L.~Gao, ``On inferring autonomous system relationships in the internet,''
  \emph{IEEE/ACM Transactions on networking}, vol.~9, no.~6, pp. 733--745,
  2001.

\bibitem{gill2013survey}
P.~Gill, M.~Schapira, and S.~Goldberg, ``A survey of interdomain routing
  policies,'' \emph{ACM SIGCOMM Computer Communication Review}, vol.~44, no.~1,
  pp. 28--34, 2013.

\bibitem{anwar2015investigating}
R.~Anwar, H.~Niaz, D.~Choffnes, {\'I}.~Cunha, P.~Gill, and E.~Katz-Bassett,
  ``Investigating interdomain routing policies in the wild,'' in
  \emph{Proceedings of the 2015 Internet Measurement Conference}, 2015, pp.
  71--77.

\bibitem{li2015route}
S.~Li, H.~Duan, Z.~Wang, and X.~Li, ``Route leaks identification by detecting
  routing loops,'' in \emph{International Conference on Security and Privacy in
  Communication Systems}.\hskip 1em plus 0.5em minus 0.4em\relax Springer,
  2015, pp. 313--329.

\bibitem{sriram2016problem}
K.~Sriram, D.~Montgomery, D.~McPherson, E.~Osterweil, and B.~Dickson, ``Problem
  definition and classification of bgp route leaks,'' \emph{RFC 7908}, 2016.

\bibitem{galmes2020preventing}
M.~F. Galm{\'e}s, R.~C. Aumatell, A.~Cabellos-Aparicio, S.~Ren, X.~Wei, and
  B.~Liu, ``Preventing route leaks using a decentralized approach: An
  experimental evaluation,'' in \emph{2020 IEEE 28th International Conference
  on Network Protocols (ICNP)}.\hskip 1em plus 0.5em minus 0.4em\relax IEEE,
  2020, pp. 1--6.

\bibitem{hepner2009defending}
C.~Hepner and E.~Zmijewski, ``Defending against bgp man-in-the-middle
  attacks,'' \emph{Talk at BlackHat}, vol. 2009, 2009.

\bibitem{routleak2015google}
D.~Madory, ``Routing leak briefly takes down google,'' \emph{Online.
  https://blogs.oracle.com/internetintelligence/routing-leak-briefly-takes-down-google},
  2015.

\bibitem{exampleRouteLeak}
A.~Siddiqui, ``Major route leak by as28548,'' \emph{Online.
  https://www.manrs.org/2021/02/major-route-leak-by-as28548-another-bgp-optimizer/},
  2021.

\bibitem{jin2019stable}
Y.~Jin, C.~Scott, A.~Dhamdhere, V.~Giotsas, A.~Krishnamurthy, and S.~Shenker,
  ``Stable and practical $\{$AS$\}$ relationship inference with problink,'' in
  \emph{16th $\{$USENIX$\}$ Symposium on Networked Systems Design and
  Implementation ($\{$NSDI$\}$ 19)}, 2019, pp. 581--598.

\bibitem{jin2020toposcope}
Z.~Jin, X.~Shi, Y.~Yang, X.~Yin, Z.~Wang, and J.~Wu, ``Toposcope: Recover as
  relationships from fragmentary observations,'' in \emph{Proceedings of the
  ACM Internet Measurement Conference}, 2020, pp. 266--280.

\bibitem{shapira2020unveiling}
T.~Shapira and Y.~Shavitt, ``Unveiling the type of relationship between
  autonomous systems using deep learning,'' in \emph{NOMS 2020-2020 IEEE/IFIP
  Network Operations and Management Symposium}.\hskip 1em plus 0.5em minus
  0.4em\relax IEEE, 2020, pp. 1--6.

\bibitem{he2020roachain}
G.~He, W.~Su, S.~Gao, J.~Yue, and S.~K. Das, ``Roachain: Securing route origin
  authorization with blockchain for inter-domain routing,'' \emph{IEEE
  Transactions on Network and Service Management}, 2020.

\bibitem{chen2020isrchain}
D.~Chen, Y.~Ba, H.~Qiu, J.~Zhu, and Q.~Wang, ``Isrchain: Achieving efficient
  interdomain secure routing with blockchain,'' \emph{Computers \& Electrical
  Engineering}, vol.~83, p. 106584, 2020.

\bibitem{sriram2017methods}
K.~Sriram, D.~Montgomery, B.~Dickson, K.~Patel, and A.~Robachevsky, ``Methods
  for detection and mitigation of bgp route leaks,''
  \emph{draft-ietf-idr-route-leak-detection-mitigation-06}, 2017.

\bibitem{abd2020bgp}
S.~Abd El~Monem, A.~Khalafallah, and S.~I. Shaheen, ``Bgp route leaks detection
  using supervised machine learning technique,'' in \emph{2020 2nd Novel
  Intelligent and Leading Emerging Sciences Conference (NILES)}.\hskip 1em plus
  0.5em minus 0.4em\relax IEEE, 2020, pp. 15--20.

\bibitem{hou2021systematic}
D.~Hou, J.~Zhang, K.~L. Man, J.~Ma, and Z.~Peng, ``A systematic literature
  review of blockchain-based federated learning: Architectures, applications
  and issues,'' in \emph{2021 2nd Information Communication Technologies
  Conference (ICTC)}.\hskip 1em plus 0.5em minus 0.4em\relax IEEE, 2021, pp.
  302--307.

\bibitem{shayan2020biscotti}
M.~Shayan, C.~Fung, C.~J. Yoon, and I.~Beschastnikh, ``Biscotti: A blockchain
  system for private and secure federated learning,'' \emph{IEEE Transactions
  on Parallel and Distributed Systems}, vol.~32, no.~7, pp. 1513--1525, 2020.

\bibitem{majeed2019flchain}
U.~Majeed and C.~S. Hong, ``Flchain: Federated learning via mec-enabled
  blockchain network,'' in \emph{2019 20th Asia-Pacific Network Operations and
  Management Symposium (APNOMS)}.\hskip 1em plus 0.5em minus 0.4em\relax IEEE,
  2019, pp. 1--4.

\bibitem{lu2019blockchain}
Y.~Lu, X.~Huang, Y.~Dai, S.~Maharjan, and Y.~Zhang, ``Blockchain and federated
  learning for privacy-preserved data sharing in industrial iot,'' \emph{IEEE
  Transactions on Industrial Informatics}, vol.~16, no.~6, pp. 4177--4186,
  2019.

\bibitem{lepinski2012rfc}
M.~Lepinski and S.~Kent, ``Rfc 6480: an infrastructure to support secure
  internet routing,'' \emph{Internet Engineering Task Force (IETF)}, 2012.

\bibitem{mcdaniel2020flexsealing}
T.~McDaniel, J.~M. Smith, and M.~Schuchard, ``Flexsealing bgp against route
  leaks: peerlock active measurement and analysis,'' \emph{arXiv preprint
  arXiv:2006.06576}, 2020.

\bibitem{jia2study}
J.~Jia, Z.-w. YAN, G.-g. GENG, and J.~Jian, ``Study on bgp route leak,''
  \emph{Chinese Journal of Network and Information Security}, vol.~2, no.~8,
  pp. 54--61, 2016.

\bibitem{siddiqui2015self}
M.~Siddiqui, D.~Montero, R.~Serral-Graci{\`a}, and M.~Yannuzzi, ``Self-reliant
  detection of route leaks in inter-domain routing,'' \emph{Computer Networks},
  vol.~82, pp. 135--155, 2015.

\bibitem{azimovverification}
A.~Azimov, E.~Bogomazov, R.~Bush, K.~Patel, and J.~Snijders, ``Verification of
  as path using the resource certificate public key infrastructure and
  autonomous system provider authorization.'' 2018.

\bibitem{yue2021privacy}
J.~Yue, Y.~Qin, S.~Gao, W.~Su, G.~He, and N.~Liu, ``A privacy-preserving route
  leak protection mechanism based on blockchain,'' in \emph{2021 IEEE
  International Conference on Information Communication and Software
  Engineering (ICICSE)}.\hskip 1em plus 0.5em minus 0.4em\relax IEEE, 2021, pp.
  264--269.

\bibitem{sabt2015trusted}
M.~Sabt, M.~Achemlal, and A.~Bouabdallah, ``Trusted execution environment: what
  it is, and what it is not,'' in \emph{2015 IEEE Trustcom/BigDataSE/ISPA},
  vol.~1.\hskip 1em plus 0.5em minus 0.4em\relax IEEE, 2015, pp. 57--64.

\bibitem{xiang2013sign}
Y.~Xiang, X.~Shi, J.~Wu, Z.~Wang, and X.~Yin, ``Sign what you really care
  about--secure bgp as-paths efficiently,'' \emph{Computer Networks}, vol.~57,
  no.~10, pp. 2250--2265, 2013.

\bibitem{zheng2018blockchain}
Z.~Zheng, S.~Xie, H.-N. Dai, X.~Chen, and H.~Wang, ``Blockchain challenges and
  opportunities: A survey,'' \emph{International Journal of Web and Grid
  Services}, vol.~14, no.~4, pp. 352--375, 2018.

\bibitem{li2020blockchain}
Y.~Li, C.~Chen, N.~Liu, H.~Huang, Z.~Zheng, and Q.~Yan, ``A blockchain-based
  decentralized federated learning framework with committee consensus,''
  \emph{IEEE Network}, vol.~35, no.~1, pp. 234--241, 2020.

\bibitem{korkmaz2020chain}
C.~Korkmaz, H.~E. Kocas, A.~Uysal, A.~Masry, O.~Ozkasap, and B.~Akgun, ``Chain
  fl: Decentralized federated machine learning via blockchain,'' in \emph{2020
  Second International Conference on Blockchain Computing and Applications
  (BCCA)}.\hskip 1em plus 0.5em minus 0.4em\relax IEEE, 2020, pp. 140--146.

\bibitem{zhang2020blockchain}
Q.~Zhang, P.~Palacharla, M.~Sekiya, J.~Suga, and T.~Katagiri, ``A blockchain
  based protocol for federated learning,'' in \emph{2020 IEEE 28th
  International Conference on Network Protocols (ICNP)}.\hskip 1em plus 0.5em
  minus 0.4em\relax IEEE, 2020, pp. 1--2.

\bibitem{weng2019deepchain}
J.~Weng, J.~Weng, J.~Zhang, M.~Li, Y.~Zhang, and W.~Luo, ``Deepchain: Auditable
  and privacy-preserving deep learning with blockchain-based incentive,''
  \emph{IEEE Transactions on Dependable and Secure Computing}, 2019.

\bibitem{chen2018machine}
X.~Chen, J.~Ji, C.~Luo, W.~Liao, and P.~Li, ``When machine learning meets
  blockchain: A decentralized, privacy-preserving and secure design,'' in
  \emph{2018 IEEE International Conference on Big Data (Big Data)}.\hskip 1em
  plus 0.5em minus 0.4em\relax IEEE, 2018, pp. 1178--1187.

\bibitem{caidaRelation}
CAIDA, ``As relationship dataset,'' \emph{Online. http://www.caida.org/data/
  as-relationships/}, 2021.

\bibitem{ketkar2017introduction}
N.~Ketkar, ``Introduction to keras,'' in \emph{Deep learning with
  Python}.\hskip 1em plus 0.5em minus 0.4em\relax Springer, 2017, pp. 97--111.

\bibitem{nair2010rectified}
V.~Nair and G.~E. Hinton, ``Rectified linear units improve restricted boltzmann
  machines,'' in \emph{Icml}, 2010.

\bibitem{kingma2014adam}
D.~P. Kingma and J.~Ba, ``Adam: A method for stochastic optimization,''
  \emph{arXiv preprint arXiv:1412.6980}, 2014.

\bibitem{mcmahan2017communication}
B.~McMahan, E.~Moore, D.~Ramage, S.~Hampson, and B.~A. y~Arcas,
  ``Communication-efficient learning of deep networks from decentralized
  data,'' in \emph{Artificial intelligence and statistics}.\hskip 1em plus
  0.5em minus 0.4em\relax PMLR, 2017, pp. 1273--1282.

\end{thebibliography}

\end{document}